\documentclass[preprint,12pt]{elsarticle}

\usepackage{amssymb}
\usepackage{multirow}
\usepackage[figurename=Fig.,tablename=Tab.]{caption}
\usepackage{subcaption}
\usepackage{url}
\usepackage{booktabs}
\usepackage{lipsum}% http://ctan.org/pkg/lipsum
\usepackage{multicol}% http://ctan.org/pkg/multicols
\usepackage{xcolor}%% use the color in the text

\usepackage{xspace}
\newcommand{\fhit}{\emph{fHit}\xspace}
\newcommand{\enn}{$e_{NN}$\xspace}
\newcommand{\bbin}{$\mathcal{B}$\xspace}

\usepackage{lineno}

\journal{Nuclear Instruments and Methods in Physics Research Section A}

\begin{document}

\begin{frontmatter}



\title{Gamma/Hadron Separation with the HAWC Observatory}


\author[IF-UNAM]{R.~Alfaro}

\author[UNACH]{C.~Alvarez}

\author[UMSNH]{J.D.~Álvarez}

\author[IF-UNAM]{J.R.~Angeles~Camacho}

\author[UMSNH]{J.C.~Arteaga-Velázquez}

\author[IF-UNAM]{D.~Avila~Rojas}

\author[PSU]{H.A.~Ayala~Solares}

\author[MTU]{R.~Babu}

\author[IF-UNAM]{E.~Belmont-Moreno}

\author[UMD]{C.~Brisbois}

\author[UNACH]{K.S.~Caballero-Mora}

\author[IA-UNAM]{T.~Capistrán\corref{cor1}}
\cortext[cor1]{Corresponding author}
\ead{tcapistran@astro.unam.mx}

\author[INAOE]{A.~Carramiñana}

\author[IFJ-PAN]{S.~Casanova}

\author[CIC-IPN]{O.~Chaparro-Amaro}

\author[UMSNH]{U.~Cotti}

\author[FCFM-BUAP]{J.~Cotzomi}

\author[INAOE]{S.~Coutiño~de~León}

\author[UdG]{E.~De~la~Fuente}

\author[UMSNH]{C.~de~León}

\author[INAOE]{R.~Diaz~Hernandez}

\author[LANL]{B.L.~Dingus}

\author[UW-Madison]{M.A.~DuVernois}

\author[LANL]{M.~Durocher}

\author[UdG]{J.C.~Díaz-Vélez}

\author[UMD]{R.W.~Ellsworth}

\author[UMD]{K.~Engel}

\author[IF-UNAM]{C.~Espinoza}

\author[UMD]{K.L.~Fan\corref{cor1}}
\ead{	klfan@terpmail.umd.edu}

\author[PSU]{M.~Fernández~Alonso}

\author[IA-UNAM]{N.~Fraija}

\author[IF-UNAM]{D.~Garcia}

\author[ITESM]{J.A.~García-González}

\author[IA-UNAM]{F.~Garfias}

\author[IA-UNAM]{M.M.~González}

\author[UMD]{J.A.~Goodman}

\author[LANL]{J.P.~Harding}

\author[IF-UNAM]{S.~Hernandez}

\author[University of Utah]{B.~Hona}

\author[MTU]{D.~Huang}

\author[UNACH]{F.~Hueyotl-Zahuantitla}

\author[MTU]{P.~Hüntemeyer}

\author[IA-UNAM]{A.~Iriarte}

\author[MPIK,CHULA,NARIT]{A.~Jardin-Blicq}

\author[ECAP]{V.~Joshi}

\author[UPP]{S.~Kaufmann}

\author[LANL]{G.J.~Kunde}

\author[IGeof-UNAM]{A.~Lara}

\author[IA-UNAM]{W.H.~Lee}

\author[UOS]{J.~Lee}

\author[IF-UNAM]{H.~León~Vargas}

\author[MSU]{J.T.~Linnemann\corref{cor1}}
\ead{	linneman@msu.edu}

\author[UPP]{G.~Luis-Raya}

\author[MSU]{J.~Lundeen}

\author[LANL]{K.~Malone}

\author[MPIK]{V.~Marandon}

\author[FCFM-BUAP]{O.~Martinez}

\author[CIC-IPN]{J.~Martínez-Castro}

\author[UNM]{J.A.~Matthews}

\author[UAEH]{P.~Miranda-Romagnoli}

\author[UMSNH]{J.A.~Morales-Soto}

\author[IFJ-PAN]{A.~Nayerhoda}

\author[ICN-UNAM]{L.~Nellen}

\author[MSU]{M.U.~Nisa}

\author[UAEH]{R.~Noriega-Papaqui}

\author[MPIK]{L.~Olivera-Nieto}

\author[Stanford]{N.~Omodei}

\author[MSU]{A.~Peisker}

\author[IA-UNAM]{Y.~Pérez~Araujo}

\author[UPP]{E.G.~Pérez-Pérez}

\author[UOS]{C.D.~Rho}

\author[INAOE]{D.~Rosa-González}

\author[MPIK]{E.~Ruiz-Velasco}

\author[FCFM-BUAP]{H.~Salazar}

\author[IFJ-PAN,CSIC]{F.~Salesa~Greus}

\author[IF-UNAM]{A.~Sandoval}

\author[UCSC,HKU,HKULab]{P.~M.~Saz~Parkinson\corref{cor1}}
\ead{pablosp@hku.hk}

\author[IF-UNAM]{J.~Serna-Franco}

\author[UMD]{A.J.~Smith}

\author[University of Utah]{R.W.~Springer}

\author[UPP]{O.~Tibolla}

\author[MSU]{K.~Tollefson}

\author[INAOE]{I.~Torres\corref{cor1}}
\ead{ibrahim@inaoep.mx}

\author[SJTU]{R.~Torres-Escobedo}

\author[MTU]{R.~Turner}

\author[INAOE]{F.~Ureña-Mena}

\author[FCFM-BUAP]{L.~Villaseñor}

\author[MTU]{X.~Wang}

\author[UOS]{I.J.~Watson}

\author[MPIK]{F.~Werner}

\author[UMD]{E.~Willox}

\author[MSFC]{J.~Wood}

\author[CINVESTAV]{A.~Zepeda}

\author[SJTU]{H.~Zhou}

\address[IF-UNAM]{Instituto de F\'{i}sica, Universidad Nacional Autónoma de México, Ciudad de Mexico, Mexico}

\address[UNACH]{Universidad Autónoma de Chiapas, Tuxtla Gutiérrez, Chiapas, México}

\address[UMSNH]{Universidad Michoacana de San Nicolás de Hidalgo, Morelia, Mexico}

\address[PSU]{Department of Physics, Pennsylvania State University, University Park, PA, USA}

\address[MTU]{Department of Physics, Michigan Technological University, Houghton, MI, USA}

\address[UMD]{Department of Physics, University of Maryland, College Park, MD, USA}

\address[IA-UNAM]{Instituto de Astronom\'{i}a, Universidad Nacional Autónoma de México, Ciudad de Mexico, Mexico}

\address[INAOE]{Instituto Nacional de Astrof\'{i}sica, Óptica y Electrónica, Puebla, Mexico}

\address[IFJ-PAN]{Institute of Nuclear Physics Polish Academy of Sciences, PL-31342 IFJ-PAN, Krakow, Poland}

\address[CIC-IPN]{Centro de Investigaci\'on en Computaci\'on, Instituto Polit\'ecnico Nacional, M\'exico City, M\'exico}

\address[FCFM-BUAP]{Facultad de Ciencias F\'{i}sico Matemáticas, Benemérita Universidad Autónoma de Puebla, Puebla, Mexico}

\address[UdG]{Departamento de F\'{i}sica, Centro Universitario de Ciencias Exactase Ingenierias, Universidad de Guadalajara, Guadalajara, Mexico}

\address[LANL]{Physics Division, Los Alamos National Laboratory, Los Alamos, NM, USA}

\address[UW-Madison]{Department of Physics, University of Wisconsin-Madison, Madison, WI, USA}

\address[ITESM]{Tecnologico de Monterrey, Escuela de Ingenier\'{i}a y Ciencias, Ave. Eugenio Garza Sada 2501, Monterrey, N.L., Mexico, 64849}

\address[University of Utah]{Department of Physics and Astronomy, University of Utah, Salt Lake City, UT, USA}

\address[MPIK]{Max-Planck Institute for Nuclear Physics, 69117 Heidelberg, Germany}

\address[CHULA]{Department of Physics, Faculty of Science, Chulalongkorn University, 254 Phayathai Road, Pathumwan, Bangkok 10330, Thailand}

\address[NARIT]{National Astronomical Research Institute of Thailand (Public Organization), Don Kaeo, MaeRim, Chiang Mai 50180, Thailand}

\address[ECAP]{Erlangen Centre for Astroparticle Physics, Friedrich-Alexander-Universit\"at Erlangen-N\"urnberg, Erlangen, Germany}

\address[UPP]{Universidad Politecnica de Pachuca, Pachuca, Hgo, Mexico}

\address[IGeof-UNAM]{Instituto de Geof\'{i}sica, Universidad Nacional Autónoma de México, Ciudad de Mexico, Mexico}

\address[UOS]{University of Seoul, Seoul, Rep. of Korea}

\address[MSU]{Department of Physics and Astronomy, Michigan State University, East Lansing, MI, USA}

\address[UNM]{Dept of Physics and Astronomy, University of New Mexico, Albuquerque, NM, USA}

\address[UAEH]{Universidad Autónoma del Estado de Hidalgo, Pachuca, Mexico}

\address[ICN-UNAM]{Instituto de Ciencias Nucleares, Universidad Nacional Autónoma de Mexico, Ciudad de Mexico, Mexico}

\address[Stanford]{Department of Physics, Stanford University: Stanford, CA 94305–4060, USA}

\address[CSIC]{Instituto de F\'{i}sica Corpuscular, CSIC, Universitat de València, E-46980, Paterna, Valencia, Spain}

 \address[UCSC]{Santa Cruz Institute for Particle Physics, Department of Physics, University of California at Santa Cruz, Santa Cruz, CA 95064, USA} 

\address[HKU]{Department of Physics, The University of Hong Kong, Pokfulam Road, Hong Kong, China} 

\address[HKULab]{Laboratory for Space Research, The University of Hong Kong, Hong Kong, China}

\address[SJTU]{Tsung-Dao Lee Institute, Shanghai Jiao Tong University, Shanghai, China}

\address[MSFC]{NASA Marshall Space Flight Center, Astrophysics Office, Huntsville, AL 35812, USA}

\address[CINVESTAV]{Departamento de Física, Centro de Investigación y de Estudios Avanzados del IPN, Ciudad de México, México}

\begin{abstract}
   The High Altitude Water Cherenkov (HAWC) gamma-ray observatory observes atmospheric showers produced by incident gamma rays and cosmic rays with energy from 300 GeV to more than 100 TeV. A crucial phase in analyzing gamma-ray sources using ground-based gamma-ray detectors like HAWC is to identify the showers produced by gamma rays or hadrons. The HAWC observatory records roughly 25,000 events per second, with hadrons representing the vast majority ($>99.9\%$) of these events.  The standard gamma/hadron separation technique in HAWC uses a simple rectangular cut involving only two parameters. This work describes the implementation of more sophisticated gamma/hadron separation techniques, via machine learning methods (boosted decision trees and neural networks), and summarizes the resulting improvements in gamma/hadron separation obtained in HAWC.
\end{abstract}




\end{frontmatter}

%%%%%%%%% = INTRODUCTION =
\section{Introduction}

Technological advances have enabled the expansion of the study of the cosmos to wavebands outside the small window in the optical region. The most energetic astrophysical sources emit radiation primarily in the gamma-ray band. One of the crucial issues in using ground-based detectors to study gamma-ray sources at Very High Energy (50~GeV~-~100~TeV) and Ultra-High Energy (100~TeV~-~100~PeV) is that the vast majority ($>99.9\%$) of air showers detected come from cosmic rays, rather than gamma rays.

Ground-based gamma-ray observatories detect the passage of secondary particles produced after a primary particle impinges on an atmospheric nucleus, leading to the generation of an Extensive Air Shower (EAS). Using ground level data, EAS properties can be characterized via a set of parameters, and then used to deduce the nature of the primary particle.
While gamma-ray induced showers contain mainly positrons, electrons, and gamma rays\footnote{Though they may contain {\it some} muons, their numbers are small.}, hadron-induced showers contain muons from the decay of secondary charged pions and kaons. These muons, typically created with high transverse momentum, result in hadronic showers being more spread out, with a multi-core structure, compared to gamma-ray-induced showers, which are more compact, with a single-core structure~\cite{easbook}.

Machine Learning Techniques (MLT) are a set of statistical and computer algorithms that can be used to build complex, non-linear, models from data, to tackle a broad range of tasks, including some in gamma-ray astronomy. On the specific task of gamma/hadron separation (hereafter simply G/H separation), ground-based gamma-ray observatories like HEGRA~\cite{westerhoff1995}, MAGIC~\cite{Albert2007yd}, H.E.S.S.~\cite{ohm09}, VERITAS~\cite{krause17}, ARGO-YBJ~\cite{Pagliaro2011}, and LHAASO-WCDA~\cite{LHAASOMLT}, among others, have reported excellent results using such techniques. 
 
 %%%%%%%%%% = HAWC Observatory =
  
\subsection{The HAWC Observatory}\label{HAWC}
 
The High-Altitude Water Cherenkov (HAWC)~\cite{Abeysekarasensitivity} gamma-ray observatory is a second-generation ground-based instrument located on the northern slope of the Sierra Negra volcano in the state of Puebla, Mexico, at an altitude of 4,100 meters above sea level. Like its predecessor, Milagro~\cite{Milagro07,Atkins2003}, HAWC is based on the water Cherenkov technique. It consists of an array of 300 water Cherenkov detectors, each made of a cylindrical metal structure, 7.3 meters in diameter and 5 meters high, containing 180,000 liters of purified water and four photomultiplier tubes (PMTs) at the bottom. The PMTs detect Cherenkov light generated by the secondary particles of the EAS as they traverse the water. The HAWC software trigger requires 28 PMT hits within a 150 ns time window, which results in roughly 25,000 events being recorded every second~\cite{ABEYSEKARA2018138}. The direction of the primary particle is reconstructed using the PMT timing information, while the shower core is computed using the charge on the PMTs. Thus, by measuring the detected charge and time at the PMTs, HAWC can reconstruct the characteristics of the EAS~\cite{Smith:2015wva}.

Because HAWC detects $>$99.9\% charged cosmic-ray (hadron) events, the level of background must be significantly reduced in order to perform gamma-ray observations with HAWC. The current method of G/H separation used by the HAWC collaboration applies a simple rectangular cut to the data, involving only two parameters. Cuts on these two parameters define a rectangular region containing, preferentially, gamma-ray events. Generally speaking, this is not an optimal classification strategy because the boundary between gamma-like and hadron-like events is not defined by the actual distribution of the two types of events. In addition, the performance of the two parameters depends on the size of the observed shower (they are more sensitive for large events), so determining their optimum combination is not straightforward. A non-linear classification method should, in principle, provide a more effective discriminator.

This paper describes the implementation of two new G/H separation methods in HAWC, using MLT; one based on Boosted Decision Trees (BDT) and another using Neural Networks (NN). The performance of the new techniques is compared with previously used HAWC cuts~\cite{Abeysekara2017,energyestimatorpaper}. 

The outline of the paper is as follows: Section~\ref{sec:variables} gives an overview of the key parameters generated from HAWC data, which are used as inputs in our G/H separation models. Section~\ref{sec:data} describes the HAWC data used in our study, both Monte Carlo (MC) simulated data, as well as real data on three astrophysical sources. Section~\ref{GHSM} describes the G/H separation models discussed in the paper, including the current (standard) methods used by HAWC, as well as our two new proposed techniques. Section~\ref{sec:building} describes how we build the different models, including details on determining the optimal cuts for each method. Section~\ref{Testing} reports the performance of the various methods, comparing them via MC and real data. We conclude, in Section~{\ref{DAC}}, with a discussion of the overall performance of the models, along with possible implications regarding the future improvements of our results.

%%%%%%%%% == G/H separation parameters ==
\section{HAWC G/H separation parameters}\label{sec:variables}

Among the many parameters generated by the HAWC experiment for each event, we considered those that could help to characterize the nature of the EAS, ultimately settling on seven, which we used as inputs in our G/H separation algorithms. These parameters broadly fall into three classes: those related to the energy of the event, those sensitive to the muon content of the shower, and those connected to the shower's lateral development, via the lateral charge distribution function.

\subsection{Energy parameters}\label{sec:energyparameters}

Two official gamma-ray energy estimators are currently used in HAWC: one based on charge density and the second using a neural network~\cite{energyestimatorpaper}. In both estimators, the HAWC data are grouped in a 2D binning scheme consisting of a {\it fraction hit} bin, {$\mathcal{B}$}, and an {\it energy} bin, {\it ebin}. The {$\mathcal{B}$} bin is defined as \fhit = nHit/nCh, where nHit is the number of PMTs activated during the event within 20 ns of the shower front, and nCh is the total number of PMTs in operation at the time. The energy bin ({\it ebin}) used in this work is given by the neural network energy estimator \enn~\cite{energyestimatorpaper}. We use ten\footnote{Note that the {$\mathcal{B}=0$} bin is currently not being used in standard HAWC analyses, as it has low sensitivity with the standard G/H classifiers. We nevertheless report on it here, to study the behavior of our machine learning algorithms over the full range.} {$\mathcal{B}$} bins and twelve quarter-decade energy bins, starting from 316 GeV (see Table ~\ref{Tab:fbin}).

%%%%%%%%% fractional bins
	\begin{table}[h]
		\begin{center}
			\caption{{\small Definition of the (10) fraction hit bins ({$\mathcal{B}$}) and (12) {\it ebin} bins; the latter represents the logarithm of the lower energy bound, $\log_{10}($\enn$/GeV)$, for each bin.}}
			\label{Tab:fbin}
			\scalebox{1.0}{\begin{tabular}{| l | l | l | }
				\hline
				{$\mathcal{B}$} & Range (\%) & {\it ebin} \\
				\hline
				0	& 	 4.4 -- 6.7 &  2.50\\
				\hline
				1	& 	 6.7 -- 10.5 &  2.75 \\
				\hline
				2	& 	10.5 -- 16.2 & 3.00 \\
				\hline
				3	& 	16.2 -- 24.7 & 3.25 \\
				\hline
				4	& 	24.7 -- 35.6 & 3.50 \\
				\hline
				5	& 	35.6 -- 48.5 & 3.75 \\
				\hline
				6	& 	48.5 -- 61.8 & 4.00 \\
				\hline
				7	& 	61.8 -- 74.0 & 4.25 \\
				\hline
				8	& 	74.0 -- 84.0 & 4.50 \\
				\hline
				9	& 	84.0 -- 100.0 & 4.75 \\
				\hline
%				\nodata & \nodata & 5.00 \\
                & & 5.00 \\
				\hline
%				\nodata & \nodata & 5.25 \\
                 & &  5.25 \\
				\hline
			\end{tabular}
			}
		\end{center}
	\end{table}
%%%%%%%%% fractional bins

%%%%%%%%% === Hadronic group ===

\subsection{Muon content parameters}

Typically, the muons present in a hadronic cascade are produced at a considerable distance from both the shower axis and one another. In the HAWC detector, these lead to strong signals in widely-separated PMTs. Two HAWC parameters can be used to try to identify them:

        \begin{itemize}
            \item {\it LIC} is the log transformation of the inverse of the {\it compactness} parameter, an empirical parameter originally developed by the Milagro Collaboration~\cite{Atkins2003}, as described in Abeysekara et al. 2017~\cite{Abeysekara2017}:
                \begin{center}
                    {\it LIC}= $\log_{10}\frac{1}{\it compactness}$ = $\log_{10} \frac{CxPE_{40}}{nHit}$,
                \end{center}
                where $CxPE_{40}$ is the charge measured in the PMT with the largest effective charge far ($>$40 m) from the shower core. When a muon passes near a PMT, the resulting charge (and, thus, {\it LIC}) will be large (see Figure~3 of Pretz et al. 2015~\cite{pretz2015}), indicating that the shower is more likely produced by a hadron. Since gamma ray showers contain few, if any, muons, they are characterized by a small {\it LIC} value.
                    
            \item {\it disMax} measures the physical distance, in meters, between the two brightest PMTs. Hadronic showers are expected to have large values of {\it disMax}, while gamma-ray showers are characterized by small values.
        \end{itemize}
\subsection{Lateral development parameters}

In gamma-ray showers, most secondary particles are generated close to the shower axis. Thus, HAWC registers their signals near this axis, with a smooth decrease with distance from the core. Three HAWC parameters can be used to describe the lateral development of the shower:

        \begin{itemize}
            \item {\it PINC} (Parameter for IdeNtifying Cosmic rays) is a parameter that quantifies the smoothness of the lateral charge distribution function (LDF) (see Figure~4 of Abeysekara et al. 2017~\cite{Abeysekara2017}). Gamma-ray showers are characterized by having PMTs with a high charge near the core, and a smoothly decreasing LDF. By contrast, hadronic showers typically contain several clumps of charges caused by widely-separated muons, thus leading to a ``wrinkled" LDF. PINC, in essence, is the $\chi^2$ of the difference between the effective log charge of each PMT hit 
            ($q_i$) and the expected mean value ($\langle q \rangle$) computed by averaging all PMTs within an annulus, 5~m in width, centered on the core of the air shower containing the PMT hit.
            \begin{center}
                $PINC= \frac{1}{N}\sum_{i=0}^{N}\frac{\left[\log_{10}(q_i) - \langle{\log_{10}(q_i)}\rangle\right]^2}{\sigma^2}$
            \end{center}
            Here $\sigma$ is the uncertainty in $q$, based on a study of gamma shower data from the Crab~\cite{Abeysekara2017}, and $N$ is the number of annuli.
                
            \item {\it LDFChi2} is the reduced chi-square obtained from fitting the LDF, with the expected shape given by the NKG function~\cite{nkgpaper}:
            
                 \begin{center}
                    $NKG = A\ \rho^{s-3}\ (1+\rho)^{s-4.5}$,\\
                \end{center}
            
            where $\rho$ is the distance from the shower axis ($r_{axis}$) at the observation level, in units of the Moli\`{e}re radius\footnote{$R_m$ = 124 m at HAWC.} ($\rho = r_{axis}/R_m$), A is the amplitude, and $s$ the  shower age. Because the charge distribution is more homogeneous in a gamma-ray shower, than a hadronic one~\cite{KrawczynskiVeritas}, the model fits better in gamma-ray events than hadronic ones. 
                
            \item {\it LDFAmp} is the logarithm of the amplitude obtained from the LDF fit. Gamma-ray and hadronic events in a given fraction hit bin {$\mathcal{B}$} are expected to have different values of {\it LDFAmp} because of differences in the lateral distributions of gamma vs. hadron events.  
        \end{itemize}

%%%%%%%%% === data set ==
\section{Data sets}\label{sec:data}

%%%%%%%%% == MC data == 
\subsection{Monte Carlo Data}\label{sec:mcdata}	

The Monte Carlo (MC) simulations of HAWC data are generated using a set of standard software packages (e.g., CORSIKA\footnote{https://www.iap.kit.edu/corsika/}, GEANT4\footnote{https://geant4.web.cern.ch}), in combination with HAWC-specific simulations that model the PMT response. CORSIKA 7.4~\cite{corsika} was used to simulate extensive air showers initiated by high energy particles in the atmosphere, using the QGSJET-II-04 and FLUKA hadronic interaction models. GEANT4~\cite{geant4reference} was used to simulate the passage of the shower particles through the HAWC detector. 
    
Nine species of primary particles were simulated: eight atomic nuclei\footnote{H, He, C, O, Ne, Mg, Si, and Fe.} (MC background), along with gamma rays (MC signal). Approximately 23 million signal and 13 million background events were generated, using a power-law energy spectrum with a spectral index of -2.0 between 5 GeV and 500 TeV, uniformly on the sky within a zenith angle below 60$^\circ$. The choice of a relatively hard spectrum results in increased statistics at higher energies at a considerable savings in computing time. For analyses which simulate the transit of a specific astrophysical source (e.g., the Crab Nebula, with a spectral index of -2.63), our simulated events must be weighted by energy and location.
The number of simulated events we used was found to be sufficient for previous studies carried out by the HAWC Collaboration, such as the application of neural networks to estimate the primary particle energy in HAWC~\citep{energyestimatorpaper}.

\subsection{Real HAWC data on astrophysical sources}
    
In order to test our classification models on real data, we selected all available HAWC data from June 2015 to December 2017 ($\sim$ 837 live days). We explored three different sources: the Crab Nebula, and the extra-galactic sources Markarian 421 and Markarian 501. 

%%%%%%%%% === Crab testing ===
\subsubsection*{Crab}		

The Crab is the remnant of the historical supernova explosion, recorded by Chinese astronomers in 1054. One of the most famous astrophysical objects\footnote{Also known as M1, the first entry in the famous catalog of astronomical objects compiled by Charles Messier in the 18th century.}, the Crab is detected across the electromagnetic spectrum~\cite{Crabbook} and its brightness and relatively steady flux at TeV energies have made it the definitive reference/calibration source for all TeV instruments.

\subsubsection*{Markarian 421 and 501}

Markarian 421 and 501 (hereafter Mrk 421 and Mrk 501) are two relatively nearby ($<$ 150 Mpc) Active Galactic Nuclei (AGN) of the {\it blazar} variety (i.e., with jets of accelerating particles pointed towards our line of sight)~\cite{surveying}. They have been known to emit at very high energy ($>$ 100 GeV) for decades, and they routinely experience outbursts during which they become even brighter than the Crab. HAWC detects them at high significance, and indeed, monitors them daily for any unusual activity~\cite{dailymonitoring}.

\subsection{Real HAWC data as background data}
A one-day random sample of real HAWC data (slightly larger than the MC background sample) is also used as background in determining the HAWC standard cuts (\ref{sec:sc}), and as an option in training background for MLT. In Section~\ref{sec:MCtesting}, we compare results using real vs. simulated background data.
	
%%%%%%%%%% = G/H separation models =
\section{G/H separation models}\label{GHSM}

The goal of the G/H separation task is to keep a majority of gamma-ray events while rejecting most hadron events. We define $\xi_{\gamma}$ as the fraction of gamma-ray events passing the G/H selection, in other words, the fraction of gamma-ray events correctly classified. Conversely, we define $\xi_{h}$ as the fraction of hadron events passing the G/H selection cut, and thus being misclassified. Thus, our aim is to achieve a gamma efficiency ($\xi_{\gamma}$) close to 1 while keeping the hadron misidentification rate ($\xi_{h}$) near 0. 

\begin{figure*}[h!]
			\centering
			{\includegraphics[width=0.96\textwidth,height=0.66\textwidth]{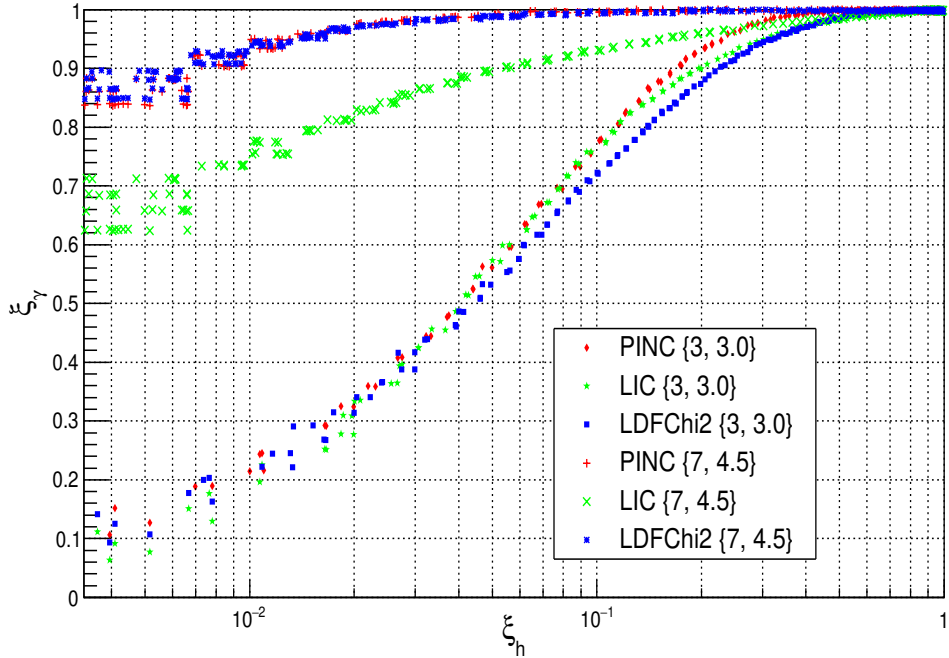}}
			\caption{{\small ROC curves of the {\it PINC} (red), {\it LIC} (green) and $LDFChi2$ (blue) parameters. These curves show the separation power of each parameter individually as a function of a cut, in two different bins; higher $\xi_{\gamma}$ at a given $\xi_{h}$ is preferred. The performance of the three parameters is better for the upper curves of the ({$\mathcal{B}=7$}, {\it ebin} 4.5), bin containing 31.6--56.2 TeV events, than the lower curves for the ({$\mathcal{B}=3$}, {\it ebin} 3.00) bin for 1.00--1.78 TeV. This reflects the fact that it is harder to discriminate gamma rays from hadrons in the low energy bins (with fewer struck PMTs) than in high energy ones.}}		
			\label{fig:variables}
		\end{figure*}

Figure~\ref{fig:variables} shows the Receiver Operating Characteristic (ROC) curves~\cite{rocpaper} for three of the shower parameters described in Section~\ref{sec:variables}. These curves, obtained from our MC simulations, illustrate the effect that varying thresholds in the different parameters have on the resulting values of $\xi_\gamma$ and $\xi_{h}$.

%%%%%%%%%%% Variables
%%%%%%%%%%% Variables

In the high energy bin (upper curves), the {\it PINC} and {\it LDFChi2} parameters have a similar response, with a good (high) $\xi_{\gamma}$ and an excellent (low) $\xi_{h}$. Both perform significantly better than {\it LIC} at high energy.  In the lower energy bin, all three parameters have roughly the same G/H performance, significantly worse than at high energy. Although {\it PINC} and {\it LDFChi2} are highly correlated (they are both based on the LDF of the gamma shower, see ~\ref{sec:intcorre}), they report different information, so we keep them both; at low energy, their performance differs more than at high energy.
 Lower {$\mathcal{B}$} bins typically have worse G/H performance because the shower has fewer PMTs participating in the event measurement.
	
In order to improve on the performance of any individual parameter, one can combine them, for example, by applying cuts on several parameters simultaneously~\cite{SumGHSepFegan}. Indeed, the current official G/H separation method in HAWC uses a simple 2 parameter cut, as described in Section~\ref{sec:sc}.

Other more sophisticated approaches include using a likelihood ratio method to combine several parameters~\citep{KrawczynskiVeritas}, or using MLT, as implemented successfully in the HEGRA~\cite{westerhoff1995} and H.E.S.S.~\cite{ohm09} observatories, among others. 

In Section \ref{sec:mlt}, we describe the implementation, in HAWC, of two new G/H separation methods using MLT, which combine the various input parameters described in Section~\ref{sec:variables}, to produce a single output value indicating the likely nature of the primary particle.

%%%%%%%%% == SC ==
\subsection{The Standard Cut (SC) in HAWC}\label{sec:sc}

Building on the experience with Milagro, where a cut on a single parameter was used successfully for G/H separation~\citep{Atkins2003}, the HAWC collaboration first implemented a similar single parameter cut, based on the {\it compactness} parameter~\citep{Abeysekarasensitivity} (as defined in Section~\ref{sec:variables}).  Subsequently, a cut on a second parameter was found to improve the performance. Rectangular cuts on these two variables as a function of the one-dimensional bins defined by $\mathcal{B}$, we refer to as the 1D standard cut (SC1D).  Similarly, the current official, or standard cut (SC), in HAWC involves selecting only events in a rectangular region defined by the same two parameters: {\it PINC} and {\it LIC} (see Section~\ref{sec:variables}), as given by the expression:
		\begin{center}
            $(LIC < C_L)~\&~(PINC < C_P)$,
		\end{center}
	    where $C_L$ and $C_P$ are the LIC and PINC parameter thresholds, respectively. 
	
	Events within this region are classified as gammas, while those outside are labeled as hadrons. The major difference between SC1D and the two-dimensional SC cut is that for SC, the  thresholds ($C_L$ and $C_P$) depend on both the fraction of PMTs activated during the event and the reconstructed primary particle energy; thus, each ({$\mathcal{B}$}, {\it ebin}) bin has a specific threshold for each parameter.

%%%%%%%%% == MLT ==
\subsection{Machine Learning Techniques}\label{sec:mlt}

In recent years, the use of computer algorithms to automatically build complex models based solely on data has been gaining ground in a range of fields, including gamma-ray astronomy. These Machine Learning Techniques (MLT) not only have the advantage of automating (and thus speeding up) repetitive tasks, but also have the potential for yielding new insights that may only be revealed as the computer processes (or ``learns" from) large quantities of data.

MLT fall under two broad categories: supervised and unsupervised. The former use ``labeled" data to train algorithms (e.g., classification), which can then be used to predict the labels/categories of new (unlabeled) data; the latter, by contrast, are applied to unlabeled data, allowing the algorithms themselves to uncover hidden structures in the data (e.g., via clustering) .  

In this work, we apply supervised learning methods to the {\it classification} task of distinguishing between gamma rays (signal) and hadrons (background). Among the large number of machine learning algorithms, we focus on two of the most successful ones: Boosted Decision Trees (BDT)~\cite{ohm09,KrawczynskiVeritas}, and Neural Networks (NN)~\cite{westerhoff1995,NNMAGIC}. We briefly describe these two algorithms, along with their inputs in the following paragraphs.

%%%%%%%%% === BDT ===
\subsubsection*{Boosted Decision Trees (BDT)}	

Traditional decision trees are a simple, non-parametric flowchart-like model, that use a series of binary sequential decision {\it nodes} to split data into {\it branches}, ultimately sorting them into {\it leaf} nodes~\citep{friedman2001elements}. They are extensively used to tackle problems of classification (e.g., signal vs. background).

Despite their advantages, simple decision trees have a number of drawbacks, including the {\it high variance problem}, where a slight change in the data can result in a significant change in the final model; in addition, a simple binary split often leads to a lack of smoothness in the model~\cite{friedman2001elements}. To overcome these problems, an ensemble of trees can be combined, to ultimately produce a more powerful, {\it boosted}, model: as more trees are added, the model ``learns" from the errors of the existing trees, and thus improves.
	
In this work, we use a Gradient boosting algorithm for our BDT model~\citep{friedman2001greedy}, as implemented in the xgboost python package\footnote{https://xgboost.readthedocs.io/en/stable/}. We use 500 trees, a low {\it learning rate}\footnote{This {\it learning rate} affects how model weights are updated, based on the estimated error at each stage.} of 0.1, to avoid large jumps around the minimum error, and a maximum tree depth of 5 nodes. For each tree, we use only a random 60\% selection for each individual tree\footnote{That is, 30\% of the total sample.}, to avoid over-fitting. The minimum value of loss reduction (error) for splitting the leaf node in each tree is set to 1. These parameters are advertised as likely to avoid overtraining.  We verified this by checking that the output distributions in testing is consistent with the training output distributions.

%%%%%%%%% === NN ===
\subsubsection*{Neural Networks (NN)}

Neural Networks (NN) are non-linear algorithms that use a collection of artificial neurons to attempt to mimic a human brain~\cite{bishopmlbook}. Artificial neurons, like their biological counterparts, are composed of {\it dendrites}, which collect input information, a {\it nucleus}, which combines and generates a signal, and finally, an {\it axon}, that sends the information to the output. The mathematical model consists of three blocks: input parameters; a synapse function, combining the input information (i.e., a sum); and an activation function defining the output, sometimes restricting it to a specific range (e.g., sigmoid, tanh, linear). Thus, NN  generally can be described as having three types of layers: an input layer, a set of hidden layers, and an output layer. The number of neurons in the input layer equals the number of input parameters. The number of hidden layers may vary, with each having any number of neurons. Typically, the neurons of the input and output layers follow a linear model (i.e., a sum as synapse function and a linear activation function, $y~=~\sum w_i~x_i$).
	
Our NN models were trained using the Toolkit for MultiVariate data Analysis (TMVA), a ROOT-integrated software package that provides a user-friendly environment for processing and evaluating MLT in high-energy physics~\cite{tmvareference}. We used a multilayer Perceptron with a 7:10:10:1 architecture\footnote{Several architectures were tested, but this one provided the best performance at a reasonable computational cost.}. The first layer has one neuron per input parameter. The two hidden layers have ten neurons each and a sigmoid activation function. Finally, the output layer has one neuron, giving the probability that an event is a gamma ray.  

%%%%%%%%% === Building the models. ===
\section{Building the models}\label{sec:building}

Both the BDT and NN models have the potential advantage over the cuts described in Section~\ref{sec:sc} of combining several number of input parameters, to produce a more powerful classifier. Ultimately, however, the effectiveness of the new classifier will depend on the discriminating power of each individual parameter, as well as the correlations among them. Seven parameters were selected as inputs for our BDT and NN algorithms, as described in Section~\ref{sec:variables}. 

In building a model based on MLT, one commonly requires three stages: training, verification, and testing~\cite{ldfbook}. The first and second stages typically work together to build the model, while the last stage is used to evaluate the performance and stability of the model. Each stage has an independent event sample; the purpose is to avoid memorizing the events instead of learning generalizable features. We chose to split our simulation data into two equal sets: 50\% for training and verification and 50\% for the testing stage. Thus, the algorithms use only half of the data to build a mathematical model that can recognize the differences between gamma-ray events and charged cosmic rays, while the remaining 50\% of the events are used to quantify the performance of the models. The output value for our models was defined in all cases as 1 for gamma-ray events, and 0 or -1 for hadrons, for the NN or BDT model, respectively.
	
Unfortunately, there is no clear answer to the question ``what is the best model?''; each has its pros and cons. Both the NN and BDT show a good performance in classification; however, their training is slow. The NN response calculation is somewhat faster than the BDT (though neither significantly affects event reconstruction time). The BDT is more robust at ignoring weak variables but is more vulnerable to overtraining. Rather than training separate models in each \{{$\mathcal{B}$} and {\it ebin}\} bin, the data were grouped into three containers and NN and BDT models were trained on these larger groups: {$\mathcal{B}=0-2$} (low), {$\mathcal{B}=3-5$} (medium) and {$\mathcal{B}=6-9$} (high). This grouping allowed us to include more training samples per model; the use of two different (albeit correlated) energy-related input parameters (see Section~\ref{sec:energyparameters}), allowed our models to better interpolate over the relatively large range of {$\mathcal{B}$} bins covered by each of these containers, as suggested in \cite{Baldi16}.

Nevertheless, the cuts applied on the model output were chosen separately for each ({$\mathcal{B}$}, {\it ebin}) pair, as described in the next section.

%%%%%%%%% = Optimizing the cut = 
\subsection*{Optimizing the cuts}\label{MLTS}

Although our models are designed for the classification task, they still allow us the freedom to choose the specific cuts that will determine the separation between the signal and background classes. In this work, we set a goal of removing as much background as possible while keeping at least 50\% of the signal. Section \ref{sec:data} describes the data set used to determine the cuts for each model. In order to define the best cut, we quantify the expected significance enhancement via the Q factor (described below). Sections \ref{sec:optimumsc} and \ref{sec:optimummlt} describe how we use this information to choose the specific cuts for the SC and MLT models, respectively; in both cases the final cuts are optimized for each individual bin.

%%%%%%%%% == Q factor ==
\subsubsection*{Q factor}\label{sec:qfactor}

The quality factor, Q, of a given selection cut is a parameter commonly used in ground-based gamma-ray astronomy (e.g., Milagro~\cite{Atkins2003}, VERITAS~\cite{KrawczynskiVeritas}) to measure the expected increase in the significance of an astrophysical source, after making the cut. Thus, optimizing the Q factor predicts the best way to classify the events. We use a Gaussian approximation to the Poisson significance improvement, assuming each bin contains a sufficiently large number of events. The Q factor is thus defined as
		\begin{equation} \label{eq:qfactor}
			Q~=~\frac{\xi _{\gamma}}{\sqrt{\xi _{h}}}.
		\end{equation}

%%%%%%%%% == SC == 
\subsection{Standard Cuts}\label{sec:optimumsc}

\begin{figure*}[h!]
		{\includegraphics[width=0.9\textwidth,height=0.7\textwidth]{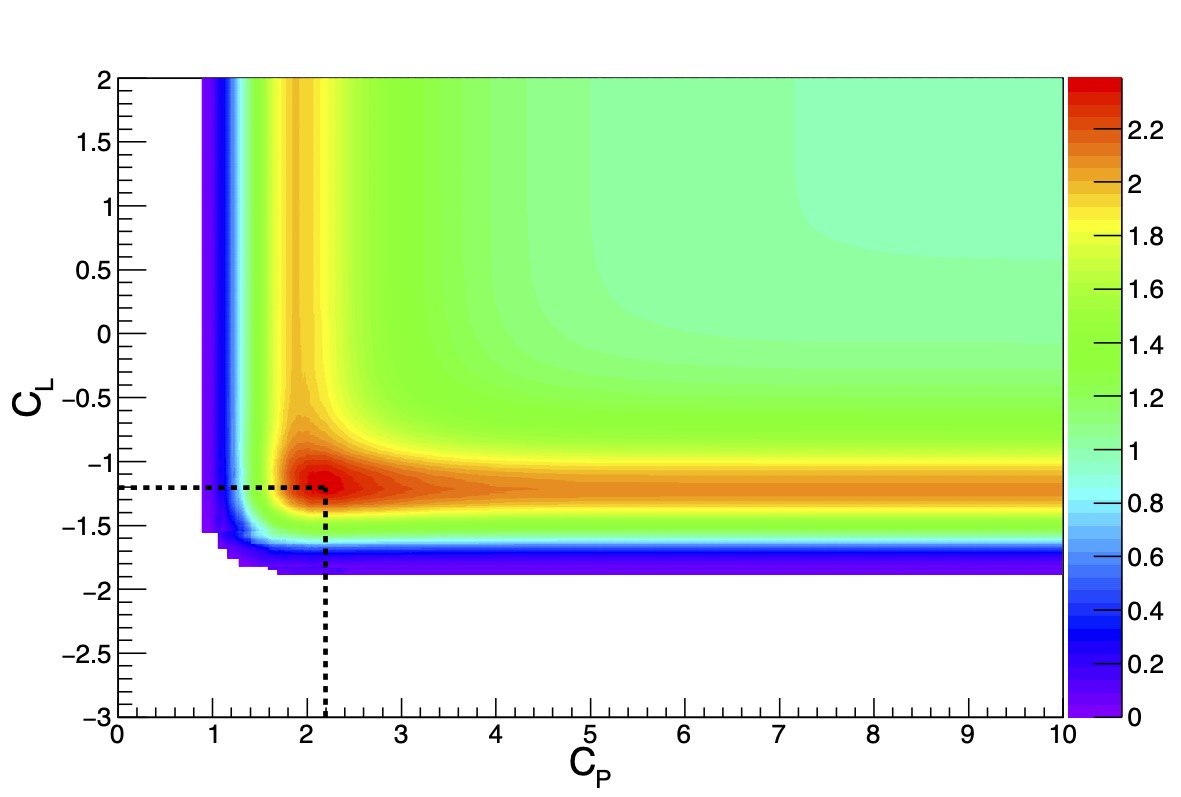}}
		\caption{{\small Q factor as a function of a cut on {\it PINC} and {\it LIC}, for ({$\mathcal{B}$}=3, {\it ebin} 3), containing 1.00--1.78 TeV. The plot illustrates the performance of the classification scheme, as a function of the chosen thresholds ($C_P$ and $C_L$). A higher Q implies a better G/H separation. The optimal cut is the point with the highest Q value. In this specific bin, this is found at $C_L=-1.202$ and $C_P= 2.195$ (indicated by the dashed lines), which retains 59.7\% of gamma-ray events, while rejecting 93.8\% of hadron events, resulting in a Q factor of 2.4. The signal region is at the lower left, enclosed by the dashed lines.}}
		\label{fig:scmodel}
	\end{figure*}

The SC involves finding optimal cuts for two parameters, separately, for each bin. First, $\xi _{\gamma}$ is computed using many candidate cuts on {\it PINC} and {\it LIC}, using the MC signal data. Next, $\xi _{h}$ is computed for these cuts using the real background set. Finally, the Q factor is calculated with Equation~\ref{eq:qfactor}, as a function of the candidate $C_P$ and $C_L$ cuts.  Figure~\ref{fig:scmodel} shows the results obtained for the ({$\mathcal{B}$}=3, {\it ebin 3.0})  bin, with energy between 1.00 and 1.78 TeV. The optimal cut values are those giving the maximum Q factor, with the proviso that at least 50\% of the gamma-ray events are retained. This process is repeated for each ({$\mathcal{B}$}, {\it ebin}) bin. Not all bin combinations contain enough data to determine the cuts, since {$\mathcal{B}$} and the particle energy are correlated; therefore, the cuts are not computed if the sample has less than 500 events.

%%%%%%%%%% finding SC
%%%%%%%%%% finding SC

%%%%%%%%% == MLT ==
\subsection{Machine Learning Techniques}\label{sec:optimummlt}

\begin{figure*}[h!]
		{\includegraphics[width=0.9\textwidth,height=0.7\textwidth]{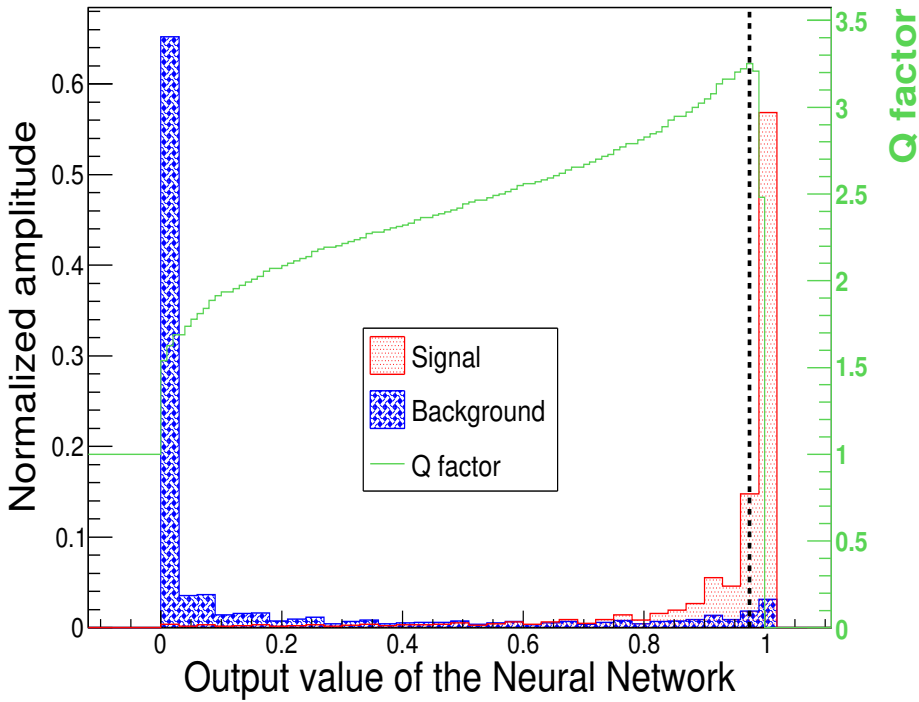}}
		\caption{{\small The probability distribution of the NN output for signal and background MC sample using the events of the {$\mathcal{B}=3$} with energy between 1.00-1.78 TeV, normalized by the number of events in each sample. The Q factor is plotted in green as a function of the cutoff on the NN output. In this specific bin, the optimal cutoff is 0.98 (dark dashed line), where it retains 63.9\% of gamma-ray events and rejects 96.1\% of hadron events, giving a maximum Q factor of 3.25.}}
		\label{fig:nnmodel}
	\end{figure*}

After the training and verification stages, the BDT and NN model outputs give the probability that an event is a gamma ray: if the output value is close to 1, there is a high probability that the event is a gamma, while an output close to 0 (or -1 for BDT), means the model predicts it is likely a background event. Figure ~\ref{fig:nnmodel} shows the distribution of the NN output using the events of the {$\mathcal{B}$}=3 bin, with energy between 1.00-1.78 TeV using signal and background MC events, as well as the corresponding Q factor as a function of threshold on the NN output. The optimal cut (0.98) for the model is the one with the maximum Q factor. As in the case of the SC, the process is repeated for each \{{$\mathcal{B}$}, {\it ebin}\} bin to find the optimal cuts for the NN and BDT models. 
	
%%%%%%%%%% finding NN cuts
%%%%%%%%%% finding NN cuts

%%%%%%%%% = Testing state ==
\section{Testing stage}\label{Testing}

The testing stage is used to evaluate and compare the models. We first test the models using samples of simulated events of known types, calculating the predicted efficiencies and Q factors (Section \ref{sec:MCtesting}). Next, we applied our G/H separation models to real data, in order to obtain the actual significances of known gamma-ray sources; specifically, we looked at three well-known sources: the Crab, Markarian 421, and Markarian 501 (Section \ref{sec:RDtesting}).

%%%%%%%%% == MC testing == 
\subsection{Testing on MC data}\label{sec:MCtesting}

Our sample of signal events was taken from the MC simulation of gamma-ray showers (see Section~\ref{sec:mcdata}), and is used in the training of all models (SC and MLT models).

For our background events, we chose two different samples; the first, from the set of background events in our MC simulation of hadron showers (see Section~\ref{sec:mcdata}). In addition, however, we used a set of randomly selected real data events (which are known to be mostly charged cosmic rays) from a single day. 

The SC model used MC signal and real data background samples for training.
The MLT models were trained on MC signal and MC background events.  The MC simulation agrees with real data (both signal and background) for all the discrimination variables~\citep{ICRC2021GHSep}. We chose to train with MC background because we obtained slightly worse MC testing results when training with real data\footnote{We also found that the NN produced significantly worse results on real Crab signals in upper \bbin bins when trained with real data.  See further discussion in \ref{sec:databkg}.}. 

Having used half of our MC sample of events for the training \& verification stages, we used the remaining half of our MC data sample for the testing stage. In order to compare the performance of all methods, we compute the Q factor for each \{{$\mathcal{B}$}, {\it ebin}\} bin for each G/H separation model, using the optimal cutoff in each case. We checked that the models were not overtrained by verifying that the model outputs on MC testing were compatible with the training outputs.  

\begin{figure*}[h!]	
			\begin{subfigure}[t]{0.49\textwidth}
				\includegraphics[width=\textwidth,height=0.9\textwidth]{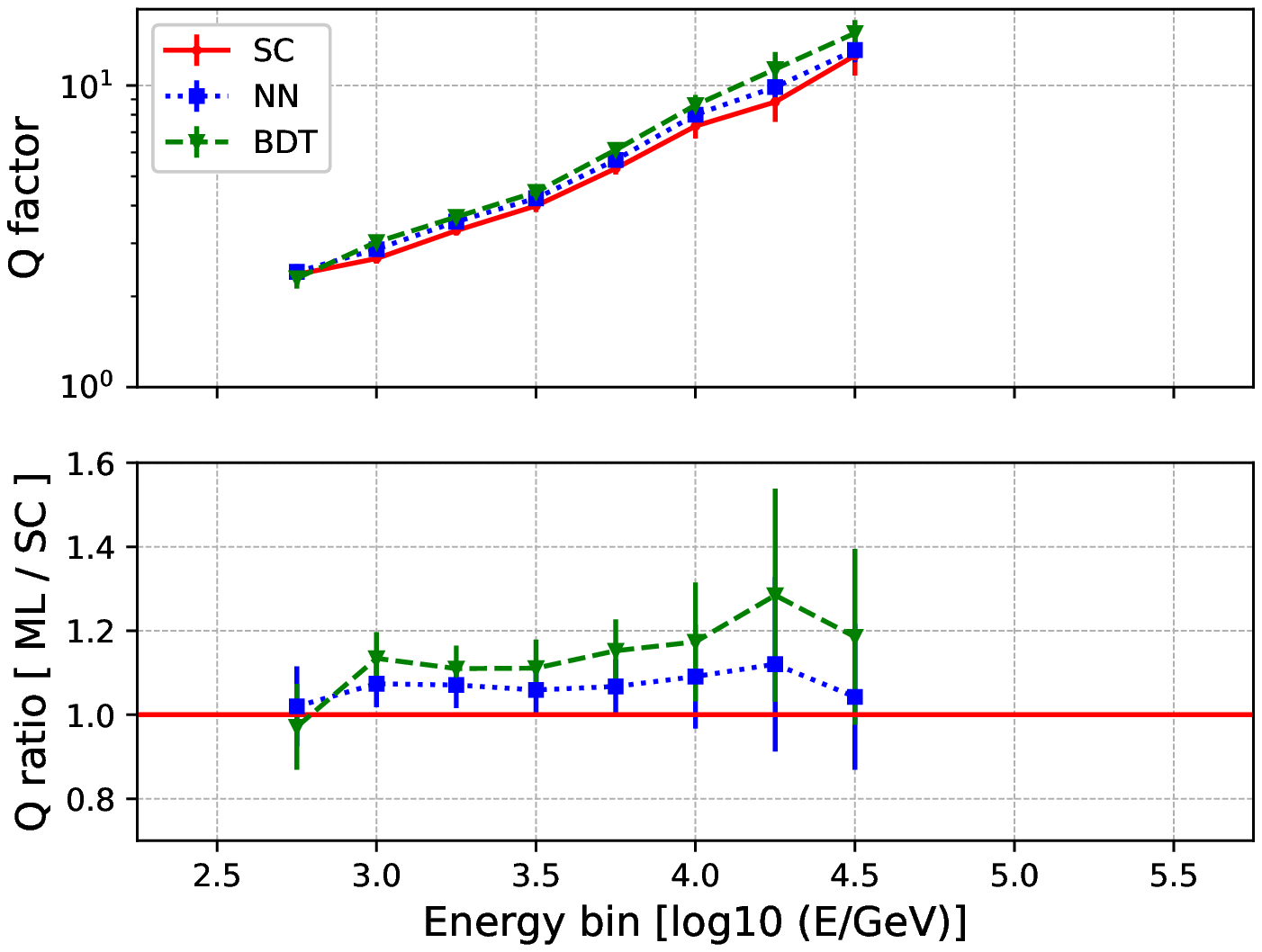}
				\caption{{\small {$\mathcal{B}=3$}}}
%				\label{#4}
			\end{subfigure}
			\begin{subfigure}[t]{0.49\textwidth}
				\includegraphics[width=\textwidth,height=0.9\textwidth]{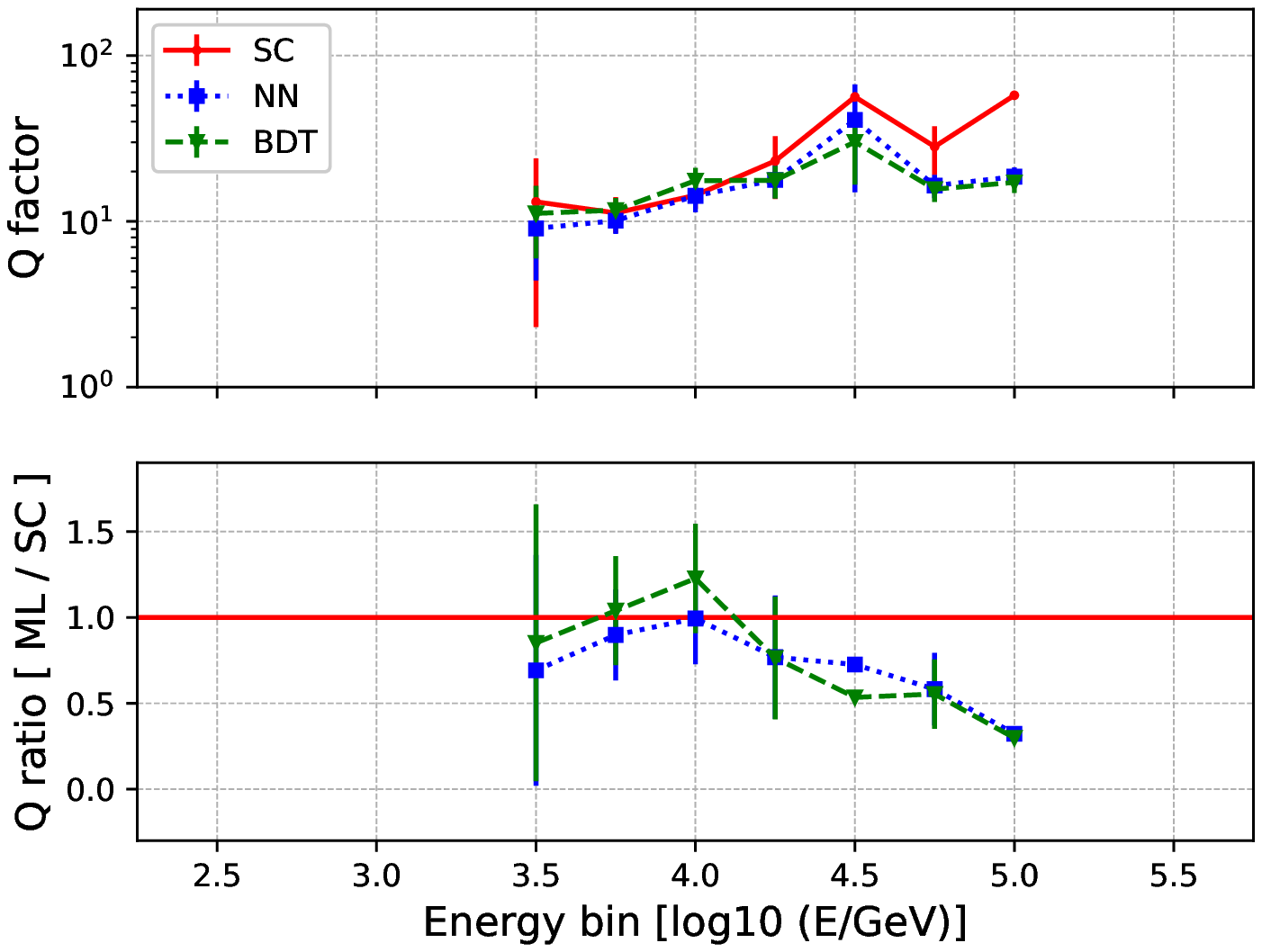}
				\caption{{\small {$\mathcal{B}=6$}}}
%				\label{#7}
			\end{subfigure}
			\caption{{\small The top panel of (a) and (b) show the Q factor for each 2D G/H separation model for the {$\mathcal{B}=3$} and {$\mathcal{B}$=6} bins, respectively, using the MC test sample. In most {\it ebins} of (a), the MLT models have better results, as reflected by the bigger Q factor, but in the case of (b), the SC shows better results at higher energies. The bottom panel of both figures shows the ratio of the Q factors for MLT models, divided by the SC. For {$\mathcal{B}=3$}, the MLT increase Q by around 10\% to 30\%.}}
			\label{fig:mctesting}
		\end{figure*}

Once we have fixed the optimal cuts for each bin, we then evaluate the predicted performance on the Crab by using the testing sample, weighted appropriately to simulate transits of the Crab.  Based on the MC results, the NN and BDT have better performance than the SC on the first six {$\mathcal{B}$} bins, while the SC is better for the rest of the bins. Figure~\ref{fig:mctesting} shows the value of the predicted Q factor of the three models for two {$\mathcal{B}$} bins (3 and 6). The bottom of the figures show the comparison of the MLT versus SC. For the {$\mathcal{B}$}=3 bin, the SC is the worst of the G/H separation models, with the NN and BDT showing an average improvement over the SC of 12\% and 30\%, respectively. On the other hand, for the {$\mathcal{B}=6$} bin, the SC reports better results than the MLT at energies above 56.2 TeV ({\it ebin}=4.75).
	
%%%%%%%%%%  MC testing
\begin{figure*}[h!]
			{\includegraphics[width=0.9\textwidth,height=0.7\textwidth]{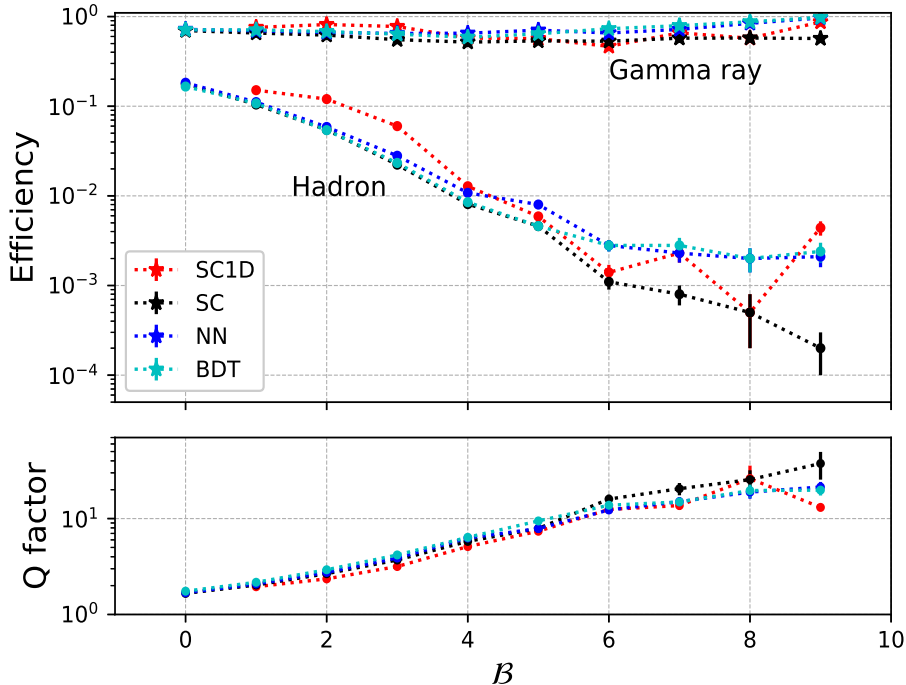}}
			\caption{{\small The gamma-ray and hadron efficiencies (top) using the MC test sample for the various classification methods: SC1D, SC, NN, and BDT. The lower panel shows the Q factor for each fit bin.}}
			\label{fig:eff}
		\end{figure*}

%%%%%%%%%% MC testing
The SC1D (see Section~\ref{sec:sc}) is the original G/H separation technique used by HAWC\footnote{Though now mostly superseded by the 2-D SC model, SC1D continues to be useful for analyses of weak or low-energy sources because it uses a less restrictive data selection than needed for applying improved energy estimators.}~\cite{Abeysekara2017}. The SC1D cuts, on PINC and compactness (and thus LiC), were optimized for each {$\mathcal{B}$} bin using a year of early Crab signal and background data.  In the initial publication, G/H separation was not attempted for $\mathcal{B}=0$. Figure~\ref{fig:eff} shows $\xi_{\gamma}$ and $\xi_{h}$ as a function of {$\mathcal{B}$} bin. The SC1D cuts were (by definition) different for each \bbin bin.  For this comparison, we applied the 2D cuts separately to each \{{$\mathcal{B}$}, {\it ebin}\}) bin, then combined the {\it ebins} belonging to each individual $\mathcal{B}$ bin.
The MLT reports a higher $\xi _{\gamma}$ at large {$\mathcal{B}$} bins. The fraction of mis-classified hadrons in the 2D models is lower in the first four {$\mathcal{B}$} bins than for SC1D, because these 2D models reject more background events. Thus, Figure~\ref{fig:eff} implies that the 2D models generally have a greater predicted Q factor, according to the MC testing comparison.
	
%%%%%%%%% MC efficiencies 
%%%%%%%%% MC efficiencies 
	
%%%%%%%%% == real testing == 
\subsection{Testing on real data}\label{sec:RDtesting}

In order to carry out tests on real data, we first applied our models to remove hadron events, and then proceeded to construct sky maps, using the official HAWC software in the standard way, as described in ~\cite{Abeysekara2017}, with a power law spectrum of index -2.7, and a pivot energy of 7 TeV.

\begin{figure*}[h!]	
			\begin{subfigure}[t]{0.49\textwidth}
				\includegraphics[width=\textwidth,height=0.9\textwidth]{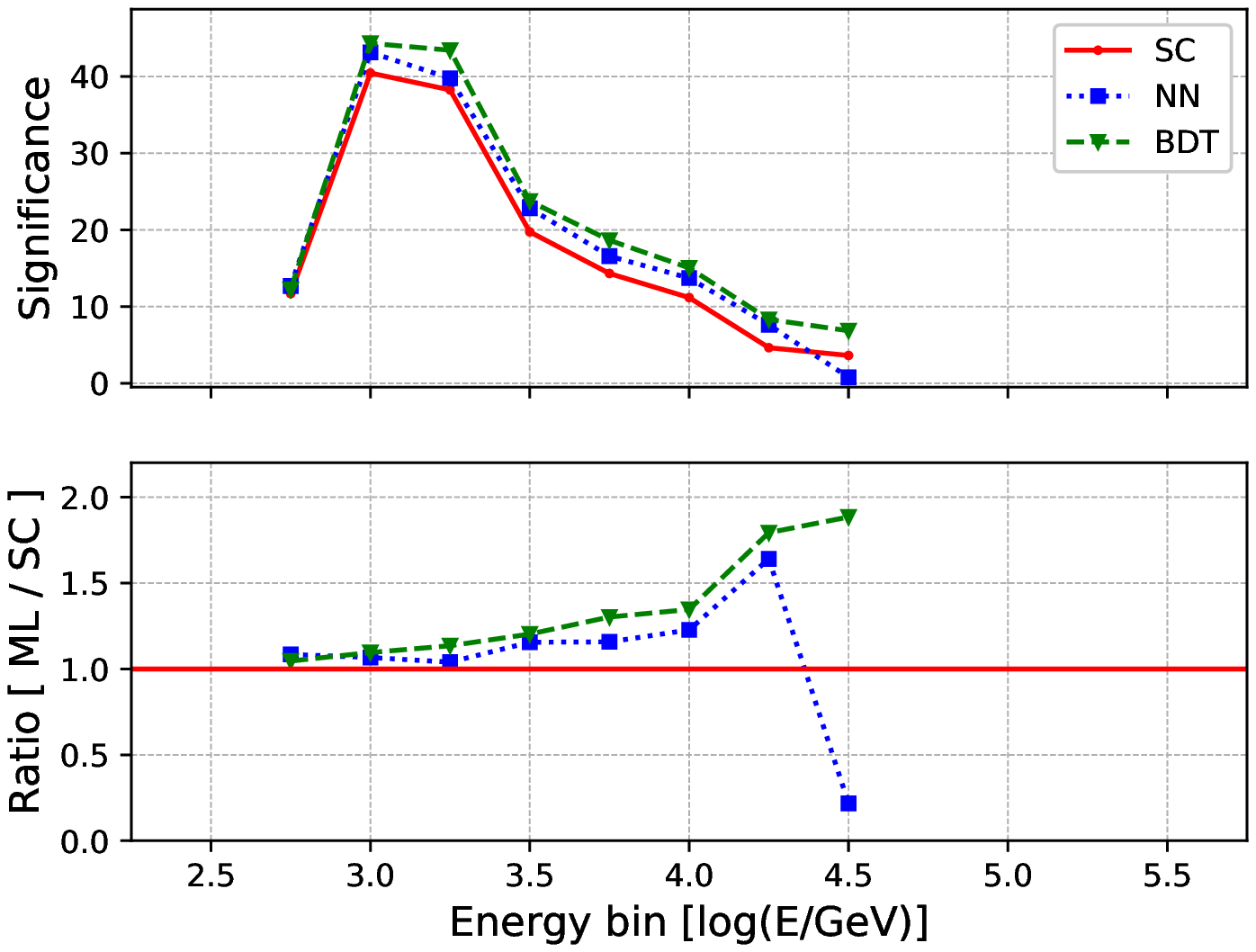}
				\caption{{\small {$\mathcal{B}$=3}}}
%				\label{#4}
			\end{subfigure}
			\begin{subfigure}[t]{0.49\textwidth}
				\includegraphics[width=\textwidth,height=0.9\textwidth]{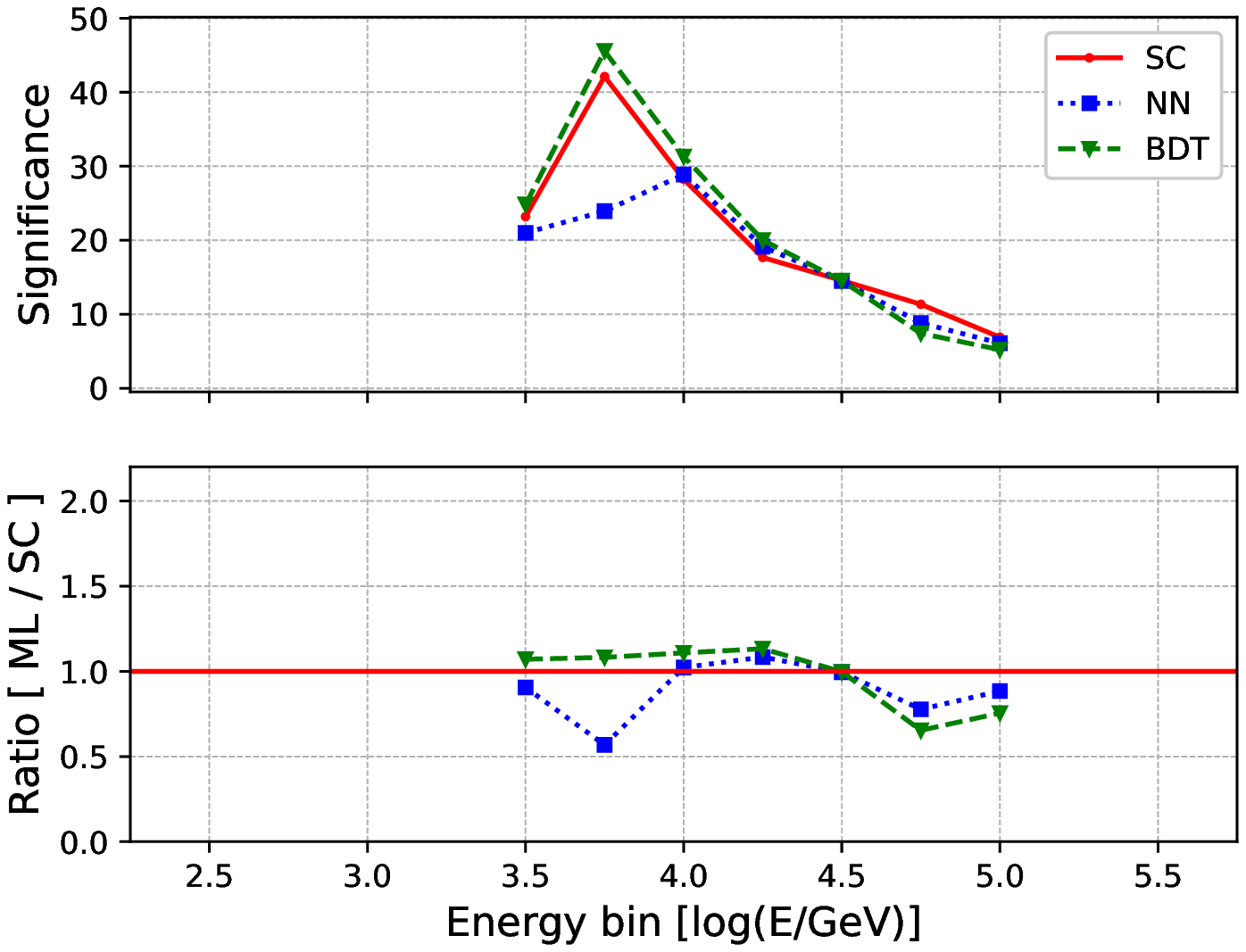}
				\caption{{\small {$\mathcal{B}=6$}}}
%				\label{#7}
			\end{subfigure}
			\caption{{\small The significance at the Crab position using the 2D models for {$\mathcal{B}=3$} (a) and {$\mathcal{B}=6$} (b) are shown in the top panel. The curves show a similar behavior to those in  Figure~\ref{fig:mctesting}, with the MLT showing a better performance than SC for {$\mathcal{B}=3$} in the most {\it ebins}, while in the {$\mathcal{B}=6$}, the results of SC are similar or higher, as can be seen from the ratio of the models, shown in the bottom panel of each figure.}}
			\label{fig:datatesting}
		\end{figure*}

The G/H separation method was used to obtain the Crab significance to show the {\it actual} performance of the various methods (rather than the predicted one, based on the MC testing set), in order to compare them. In this analysis, 67 2D bins with a significance at the source position of $>3 \sigma$ are used\footnote{Of these, four bins belong to the {$\mathcal{B}$=0}.}. For the rest of the bins (53), the maps are not included because they have too few counts or are dominated by background so that the signal is overshadowed by the noise~\cite{energyestimatorpaper}. Figure~\ref{fig:datatesting} shows the results for the {$\mathcal{B}$=3} and {$\mathcal{B}$=6} bins of the 2D G/H separation models. In the specific case of  {$\mathcal{B}$=3}, the results follow the same behavior as the testing with simulation; the MLTs show an improvement over the SC. However, in the case of  {$\mathcal{B}$=6}, the models have similar results except for energies greater than 56.2 TeV ({\it ebin} 4.75), where the SC is better. 	

%%  Crab testing
%% Crab testing

In order to determine the significance as a function of the {$\mathcal{B}$} bin, we combine all {\it ebins}, thus summarizing the performance of each G/H separation model per bin. Table~\ref{Tab:Crabfbin} reports the significance at the Crab location for each G/H separation method; the next three columns contain the fractional significance improvement of the 2D G/H separation models over the older SC1D; and the last two columns show the comparison between MLT and SC cuts. The last two rows report the combined significance using all 67 bins ({$\mathcal{B}=0-9$}), and the official bins only ({$\mathcal{B}=1-9$}). For most bins, the 2D models provide better results than SC1D. BDT improves the Crab significance compared to SC1D by 19\% for the official bins, while the SC and NN improve, by 9\% and 8\%, respectively. The BDT improves over SC in every $\mathcal{B}$ bin, while the NN improves in over half. Adding {$\mathcal{B}=0$} gives only a slight improvement, even with MLT methods, suggesting that this low bin requires a different approach if a useful signal is to be extracted from it.

%%%%%%%%% Crab table B bins with comparison between SC1D and SC 
    \begin{table*}[h]
	\begin{center}
	\caption{{\small Crab significance using each G/H separation method. Three columns show the difference, in \%, of the significances between the 2D Models and the SC1D cuts ($\frac{2D_{Model}-SC1D}{SC1D}$). The last two columns show the improvement of the MLT models over the SC cuts. The last two rows show the results from merging maps that belong to the $\mathcal{B}$ bins 1--9 and 0--9.}}
	\label{Tab:Crabfbin}
	\scalebox{0.9}{
		\begin{tabular}{c|cccc|ccc|cc}
			\hline
			\multirow{4}{*}{$\mathcal{B}$}&\multicolumn{4}{c|}{Significance} & \multicolumn{5}{c}{Difference in \% between}\\
			& & &  &  & SC & NN & BDT & NN & BDT \\ 
			& SC1D & SC &  NN & BDT & \& & \& & \&  & \& & \& \\ 
			& & &  &  & SC1D & SC1D & SC1D & SC & SC \\ 
			\hline
			0 & -  & 15.2 & 14.7 & 16.0 & -  & - & - & -3 & 5 \\ 
			1 & 26.9 & 27.6 & 27.5 & 28.22 & 3 & 2 & 5 & 0 & 2\\ 
			2 & 37.8 & 44.1 & 44.6 & 46.4 & 17 & 18 & 23 & 1 & 5\\ 
			3 & 59.2 & 62.4 & 66.1 & 72.0 & 5 & 12 & 22 & 6 & 15\\ 
			4 & 70.6 & 69.7 & 76.3 & 76.2 & -1 & 8 & 8 & 10 & 9\\ 
			5 & 67.3 & 71.3 & 69.7 & 80.1 & 6 & 4 & 19 & -2 & 12\\ 
			6 & 52.3 & 61.5 & 48.3 & 66.0 & 18 & -8 & 26 & -21 & 7\\ 
			7 & 39.1 & 47.7 & 49.2 & 50.3 & 22 & 26 & 28 & 3 & 5\\ 
			8 & 27.6 & 32.8 & 35.1 & 34.8 & 19 & 27 & 26 & 7 & 6\\ 
			9 & 28.2 & 28.7 & 31.3 & 31.3 & 2 & 11 & 11 & 9 & 9\\ 
			\hline
			1-9 & 144.0 & 155.7 & 156.9 & 170.7 & 8 & 9 & 19 & 1 & 10\\ 
			0-9 & - & 156.3 & 157.5 & 171.3 & - & - & - & 1 & 10\\ 
			\hline
   		\end{tabular}
		}
	\end{center}
    \end{table*}
%%%%%%%%% Crab table B bins with comparison between SC1D and SC
		
\begin{figure*}[h!]
				\centering
				{\includegraphics[width=0.8\textwidth,height=0.55\textwidth]{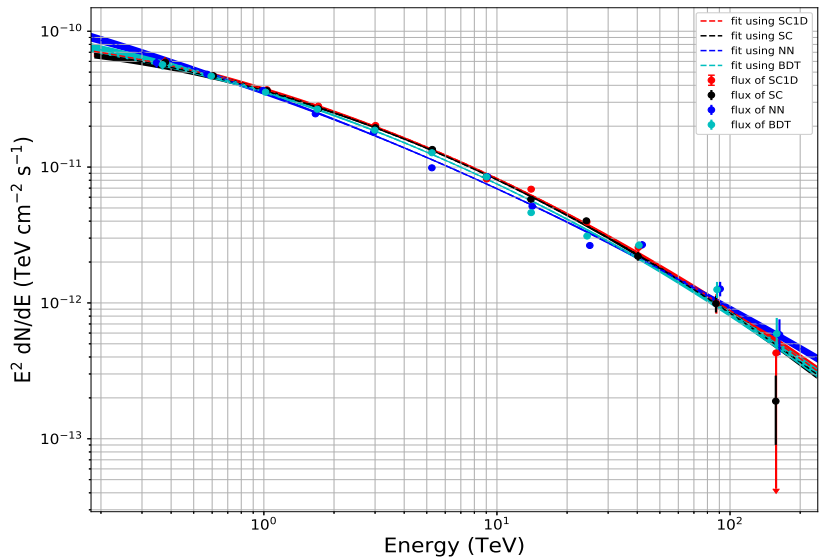}}
				\caption{{\small The Crab spectrum obtained with the SC1D (red), SC (black), NN (dark blue), and BDT (light blue) using the same method described in Abeysekara et al.~\cite{energyestimatorpaper}. The dashed lines show the spectral model fit with a log-parabola for each G/H model.}}
				\label{fig:spectrum}
			\end{figure*}

We also summarize the Crab performance as a function of the energy ({\it ebin}). The flux points were obtained for the Crab in quarter-decade energy bins, using the method described in~\cite{energyestimatorpaper}. We repeated it for each G/H separation model, using a log-parabola model to fit the spectrum (see Figure~\ref{fig:spectrum}). Table~\ref{Tab:Crabebin} reports our results, which are similar to the $\mathcal{B}$ bin projection. The 2D models give the best G/H separation in most bins. MLT gives better results than SC at low energies, but above 41.6 TeV ({\it ebin}=4.50), the SC generally has better performance.
		
%%%%%%%%% Crab spectrum
%%%%%%%%% Crab spectrum

%%%%%%%%% Crab ebin significance
\begin{table}[h]
	\begin{center}
	\caption{{\small Crab significance using each G/H separation method for the energy bin ({\it ebin}). The first column gives the lower bound for each bin ($\log($ \enn $/GeV)$).}}
	\label{Tab:Crabebin}
	\scalebox{1}{
		\begin{tabular}{c|cccc}
			\hline
			\multirow{2}{*}{\it ebin}  & \multicolumn{4}{c}{Significance}\\
			& SC1D & SC & NN & BDT \\ 
			\hline
			2.50 & 12.1 & 12.4 & 12.3 & 12.6 \\ 
			2.75 & 31.2 & 32.5 & 34.0 & 34.6 \\ 
			3.00 & 52.2 & 54.7 & 56.9 & 58.4 \\ 
			3.25 & 64.4 & 65.3 & 65.6 & 72.9  \\ 
			3.50 & 70.1 & 71.1 & 74.0 & 79.5 \\ 
			3.75 & 60.3 & 66.5 & 58.6 & 74.6 \\ 
			4.00 & 46.2 & 54.6 & 59.0 & 62.3  \\ 
			4.25 & 36.3 & 41.5 & 45.0 & 44.3 \\ 
			4.50 & 26.7 & 36.0 & 30.6 & 32.9  \\ 
			4.75 & 15.7 & 21.8 & 23.0 & 21.5  \\ 
			5.00 & 8.4 & 13.9 & 11.3 & 10.1\\ 
			5.25 & 1.9 & 3.0 & 4.8 & 4.4 \\ 
			\hline
   		\end{tabular}
		}
	\end{center}
\end{table}
%%%%%%%%% Crab ebin significance
		
%%%%%%%%% ===  Mrks testing ===

Table~\ref{Tab:mrk421fbin} and ~\ref{Tab:mrk501fbin} report the significance for Mrk 421 and Mrk 501 for each {$\mathcal{B}$} bin and for the combination of all bins (0--9 and 1--9). The MLT results for Mrk 421 are consistent with those seen in the Crab in bins where both are significantly detected. MLT has similar improvement over SC for 421 as for the Crab, but all 2D methods have smaller fractional improvement over SC1D than for the Crab. However, for Mrk 501 the NN results are worse than for SC or SC1D. The performance of the SC is better than SC1D (though again not as much as for the Crab), while the BDT improvement over SC on this source is comparable to that seen for the Crab analysis.  It is difficult to assess trends by bin for Mrk 501, because the source is not as strongly detected as Mrk 421 or the Crab.
 
%%%%%%%%% Mrk 421 table B bins with comparison between SC1D and SC 
\begin{table*}[h]
	\begin{center}
	\caption{{\small Similar to {\bf Tab.~\ref{Tab:Crabfbin}} but for Mrk 421.}}
	\label{Tab:mrk421fbin}
	\scalebox{0.9}{
		\begin{tabular}{c|cccc|ccc|cc}
			\hline
			\multirow{4}{*}{$\mathcal{B}$}&\multicolumn{4}{c|}{Significance} & \multicolumn{5}{c}{Difference in \% between}\\
			& & &  &  & SC & NN & BDT & NN & BDT\\ 
			& SC1D & SC &  NN & BDT & \& & \& & \& & \& & \&\\ 
			& & &  &  & SC1D & SC1D & SC1D & SC & SC\\ 
			\hline
			0 & - & 8.46 & 8.28 & 8.40 & - & - & - & -2 & -1\\ 
			1 & 11.9 & 13.2 & 12.5 & 13.0 & 11 & 5 & 10 & -5 & -1\\ 
			2 & 16.2 & 16.2 & 15.6 & 16.6 & 0 & -4 & 2 & -3 & 2\\ 
			3 & 19.0 & 18.9 & 19.9 & 21.2 & -1 & 4 & 11 & 5 & 12\\ 
			4 & 21.6 & 19.5 & 21.9 & 20.7 & -10 & 2 & -4 & 12 & 6\\ 
			5 & 16.5 & 15.0 & 15.5 & 17.6 & -9 & -6 & 7 & 4 & 18\\ 
			6 & 9.7 & 9.3 & 8.4 & 11.0 & -4 & -13 & 13 & -9 & 18\\ 
			7 & 4.2 & 5.6 & 7.2 & 6.9 & 34 & 72 & 65 & 28 & 23\\ 
			8 &  - & - & - & - & - & - & - & - & - \\ 
			9 &  - & - & - & - & - & - & - & - & - \\ 
			\hline
			1-9 & 35.9 & 35.3 & 36.0 & 38.6 & -2 & 0 & 8 & 2 & 10\\ 
			0-9 & - & 36.0 & 36.6 & 39.3 & - & - & - & 2 & 9\\      
			\hline
			\multicolumn{9}{c}{Crab Improvements}\\
			1-9 &  &  &  &  & 8 & 9 & 19 & 1 & 10\\ 
			\hline
		\end{tabular}
		}
	\end{center}
\end{table*}
%%%%%%%%% Mrk 421 table B bins with comparison between SC1D and SC 

%%%%%%%%% Mrk 501 table B bins with comparison between SC1D and SC 
\begin{table*}[h]
	\begin{center}
	\caption{{\small Similar to {\bf Tab.~\ref{Tab:Crabfbin}} but for Mrk 501.}}
	\label{Tab:mrk501fbin}
	\scalebox{0.9}{
		\begin{tabular}{c|cccc|ccc|cc}
			\hline
			\multirow{4}{*}{$\mathcal{B}$}&\multicolumn{4}{c|}{Significance} & \multicolumn{5}{c}{Difference in \% between}\\
			& & &  &  & SC & NN & BDT & NN & BDT\\ 
			& SC1D & SC &  NN & BDT & \& & \& & \& & \& & \&\\ 
			& & &  &  & SC1D & SC1D & SC1D & SC & SC\\ 
			\hline
			0 &  - & - & - & - & - & - & - & -  & - \\ 
			1 & 3.4 & 3.8 & 4.2 & 4.6 & 12 & 25 & 36 & 11 & 21\\ 
			2 & 4.5 & 2.9 & 3.1 & 3.7 & -36 & -32 & -17 & 6 & 29\\ 
			3 & 4.7 & 5.3 & 4.5 & 4.2 & 14 & -5 & -10 & -16 & -21\\ 
			4 & 5.1 & 5.1 & 6.2 & 4.4 & 0 & 20 & -14 & 20 & -14\\ 
			5 & 4.1 & 3.8 & 4.3 & 5.7 & -9 & 4 & 38 & 15 & 51\\ 
			6 & 3.8 & 5.0 & 2.0 & 5.7 & 31 & -47 & 50 & -59 & 14\\ 
			7 & 1.6 & 2.2 & 2.5 & 2.9 & 43 & 60 & 85 & 12 & 30\\ 
			8 & 2.6 & 2.7 & 2.3 & 2.9 & 3 & -10 & 12 & -13 & 8\\ 
			9 &  - & - & - & - & - & - & - & -  & - \\ 
			\hline
			1-9 & 10.3 & 10.6 & 10.2 & 11.9 & 4 & 0 & 16 & -4 & 12\\ 
			\hline
			\multicolumn{9}{c}{Crab Improvements}\\
			1-9 &  &  &  &  & 8 & 9 & 19 & 1 & 10\\ 
			\hline

		\end{tabular}
		}
	\end{center}
\end{table*}
%%%%%%%%% Mrk 501 table B bins with comparison between SC1D and SC 

%%%%%%%%% = D&C =
\section{Discussion and Conclusions}\label{DAC}

The current G/H separation method used by HAWC is based on a simple rectangular cut involving only two parameters. However, the sensitivity of high energy observatories depends strongly on their ability to reject hadrons, because these overshadow the gamma-ray signal coming from astrophysical sources by several orders of magnitude. To improve on the performance of current methods, we must combine the information of additional parameters. We investigated new methods using MLT to improve the G/H separation over the official standard cuts (SC and SC1D). We focus on two techniques, Neural Networks (NN) and Boosted Decision Trees (BDT), which have proven to be highly effective in a range of applications (including in VHE gamma-ray astronomy~\cite{krause17,ohm09}).
		
The machine learning models were trained and tested on the standard HAWC MC data, simulating an astrophysical source with energy spectrum and declination similar to the Crab. These methods were compared, using simulated data, with the HAWC official cuts (SC1D and SC, see Figure~\ref{fig:eff}), with the MLT models resulting in a hadron rejection similar to the SC for low {$\mathcal{B}$} bins, but a higher $\xi_{\gamma}$ at high {$\mathcal{B}$} bins.
	
We then tested the models using real data. From figure~\ref{fig:mctesting}, MC predicts that NN and BDT models have a greater Q factor than SC in the {$\mathcal{B}=3$} bin, and this is borne out in practice, based on the observed significance for the Crab (using real HAWC data) presented in Figure~\ref{fig:datatesting}. Similarly, for the {$\mathcal{B}=6$} bin, SC has a better performance in the high-energy bin ({\it ebin}). 

A summary of our Crab results is shown in Tables~\ref{Tab:Crabfbin} and \ref{Tab:Crabebin}, where it is clear that all the 2D models have better performance than SC1D (cuts binned in \bbin only). This is of interest because SC1D was tuned on Crab data and real background, while SC and MLT use MC signal.  The BDT is the best overall G/H separation model, with an improvement of $\sim10$\% over the best-present-practice SC and $\sim19$\% over SC1D.  While BDT improves over SC in all \bbin bins, the improvements were not as prominent in the higher {\it ebins} as in the lower bins, perhaps because of limited MC statistics at high energy or residual simulation modeling issues.  All of the 2D models would have benefited from larger background samples for tuning the bin cuts, as in some upper bins fewer than 100 background events passed the cuts. It is worth noting that the MLT models had the SC variables as inputs but were unable to improve on SC in most high-energy {\it ebins}.  
		
The models were also applied to two additional astrophysical gamma-ray sources: Mrk 421 and Mrk 501, two well-known extra-galactic objects with different energy spectra and declination than the Crab, for which all cuts had been tuned. The BDT gave an excellent performance in most {$\mathcal{B}$} bins, and the overall improvement in \bbin (1-9) with respect SC1D is 8\% and 16\% on Mrk 421 and 501, respectively. 
The NN had similar performance to SC1D on the two Markarians, while the 2-dimensional standard cut (SC) only slightly improved over SC1D (by less than one sigma) in Mrk 501 and was worse for Mrk 421.
This may be due to the differences in source declination or energy spectrum, compared to the Crab, which extends to higher energy and transits nearly overhead at HAWC.  But in the case of SC, it also could reflect some differences between using real Crab photon signal for SC1D and the MC photon signal used in tuning SC (and MLT).

The BDT consistently improved the observed significance over present state of the art SC by 10\%, 10\%, and 12\% for the Crab, Mrk 421, and Mrk 501, respectively.  The NN results reflect less of an improvement over SC: 1\%, 2\%, and -4\% respectively.  The BDT does not seem to be strongly dependent on the differences in the strength, declination, or spectra of the sources.  However, for most present HAWC analyses, the gains shown by the BDT are not felt to be large enough to be worth adding  the corresponding additional systematic uncertainty.

General experience in the High Energy Physics (HEP) community has been that BDT often outperforms neural nets. BDT is also typically more robust to weak or correlated variables, because of the algorithm's explicit focus on incremental variable selection.  A significant part of BDT's advantage may be simply having more free parameters. 
The neural network energy estimator~\citep{energyestimatorpaper} has 479 parameters, while the 3 NN models together have 670 parameters. The SC works with 134 parameters and the BDT, with 1500 trees, has up to 90K parameters.  Because of lower weights on later trees and the automated leaf pruning, the effective number of parameters might be considerably lower, but the BDT has at least an order of magnitude more parameters than the NN. Despite its larger size, the BDT  generalized better from the training sample than the NN, so it is unlikely that the MC sample size intrinsically limited the smaller NN model. But larger background samples (particularly at high energy) might well have further improved the bin-by-bin cut optimization and performance of MLT, and possibly of the SC as well.

The MLT are powerful algorithms that help to improve the recognition between gamma rays and hadrons. In this paper, we show an improvement in three known sources. However, the performance of these models in other sources with different characteristics (e.g. those reported in the third HAWC catalog~\cite{3hwc}) is yet to be determined. On the other hand, the field of MLT is vast, and includes many more models than the ones explored here. For example, Convolutional Neural Networks could be explored that can be trained with weakly supervised learning~\cite{withoutlabels}, where the primary goal would be to build a model with pure Crab data that avoids the discrepancy between training and testing data~\cite{WatsonICRC2021}.
		
%appendix

%++++++++++++++++++++++++++++++++++++++++++++++++++++++++++++++++++

\appendix 
 
 \section{MC vs. Data Background}\label{sec:databkg}
A surprise in our study was that training MC signal against MC background produced better results than training against our real data background sample. This is despite the real data sample having more events, including in the highest energy bins.  One would expect to do better with real background.  In general we had slightly better results in MC testing when using MC background, for both NN and BDT.  But on real Crab data, the NN performance was significantly worse in the top \bbin bins using event data background.  However, the BDT Crab results were similar when trained with either background. We looked into various possible explanations.

One might wonder whether this could be caused by problems in correctly simulating the distributions of discriminating variables.  We had studied these variables before beginning training of the models, and published results \citep{ICRC2021GHSep} showing that we saw no significant problems with the simulation matching data compared to real data around the Crab nebula, at least until upper bins where real data necessarily runs out of statistics. Our comparisons included both a background region, and a background-subtracted signal region.  Further, one would have expected both ML methods to be similarly affected by any MC vs data discrepancy. 

Adding the interpolation energy variables \fhit and \enn improved MC testing results by a few \%. While we had been thinking of them as interpolation variables, the MLT can treat them as discriminating variables.  The upper tail of the \fhit distribution (the highest \bbin bins), while similar between MC signal and MC background, differed between MC background and data background.  This reflects differences in the number of available PMTs in simulation compared to data. The MC attempted to sample appropriately over long-term detector evolution, while we used only a single data run to form the MLT training data background sample. Again, one would have naively expected this to affect BDT and NN similarly, but we believe it affected NN more (see \ref{sec:intcorre}).

In the original ML interpolation publication\citep{Baldi16}, the interpolation was on a signal theory parameter, with the background (randomly) forced to have exactly the same distribution.  Using measured values, we could not force the distributions to be identical and restrict the energy variables to interpolation, leading to some sensitivity to the distributions of the interpolation variables.  However, the choice to train with MC background added some robustness, since signal and background were generated with the same PMT availability.  Using data as background requires care to ensure a compatible detector setup between the data selected, and that in the signal MC.

\section{Correlation and Variable Importance Effects}\label{sec:intcorre}

\begin{figure*}[h!]
			\begin{subfigure}[t]{0.49\textwidth}
		    	\includegraphics[width=\textwidth,height=0.9\textwidth]{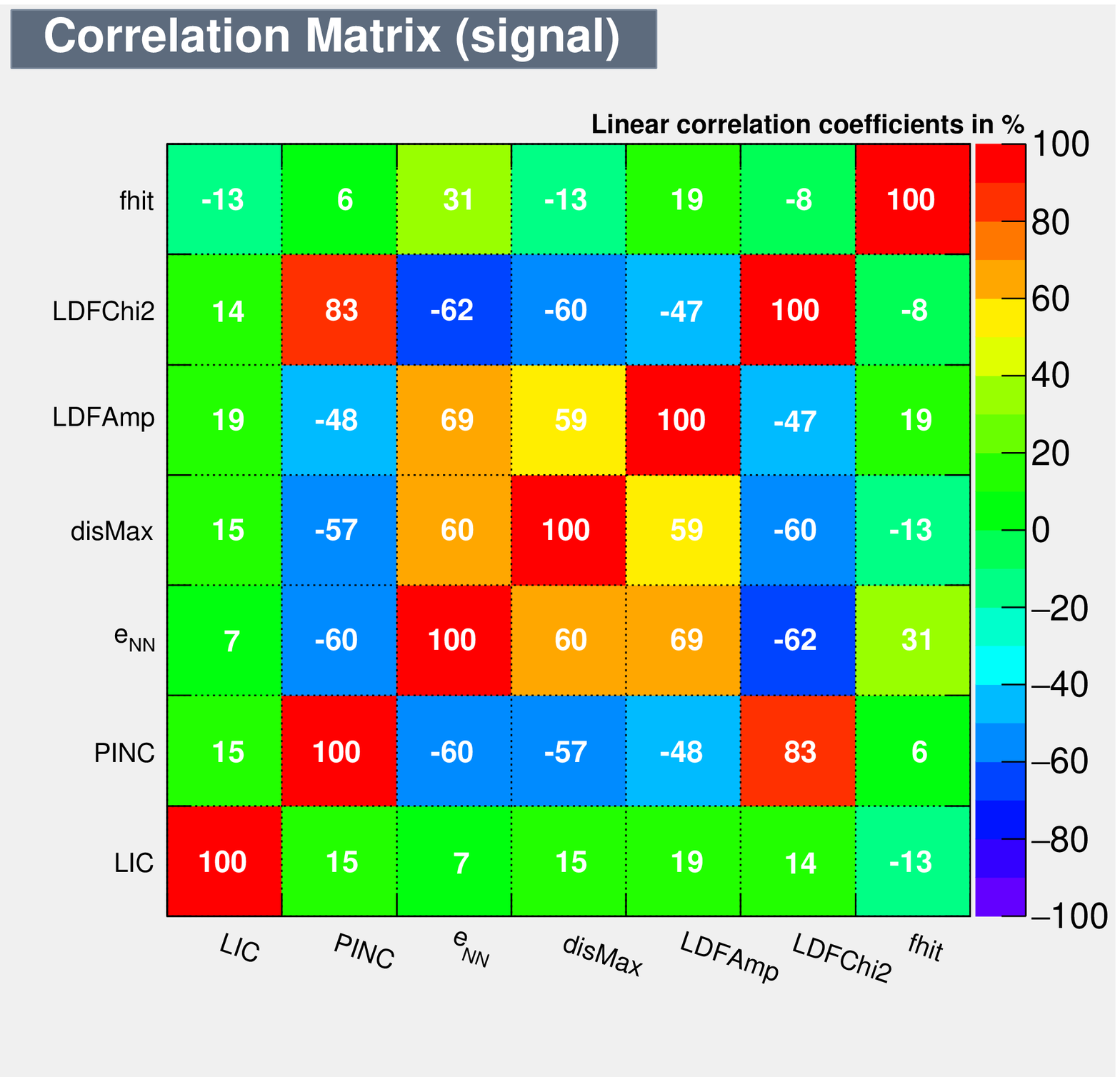}
				\caption{{\small Signal.}}
%				\label{#4}
			\end{subfigure}
			\begin{subfigure}[t]{0.49\textwidth}
				\includegraphics[width=\textwidth,height=0.9\textwidth]{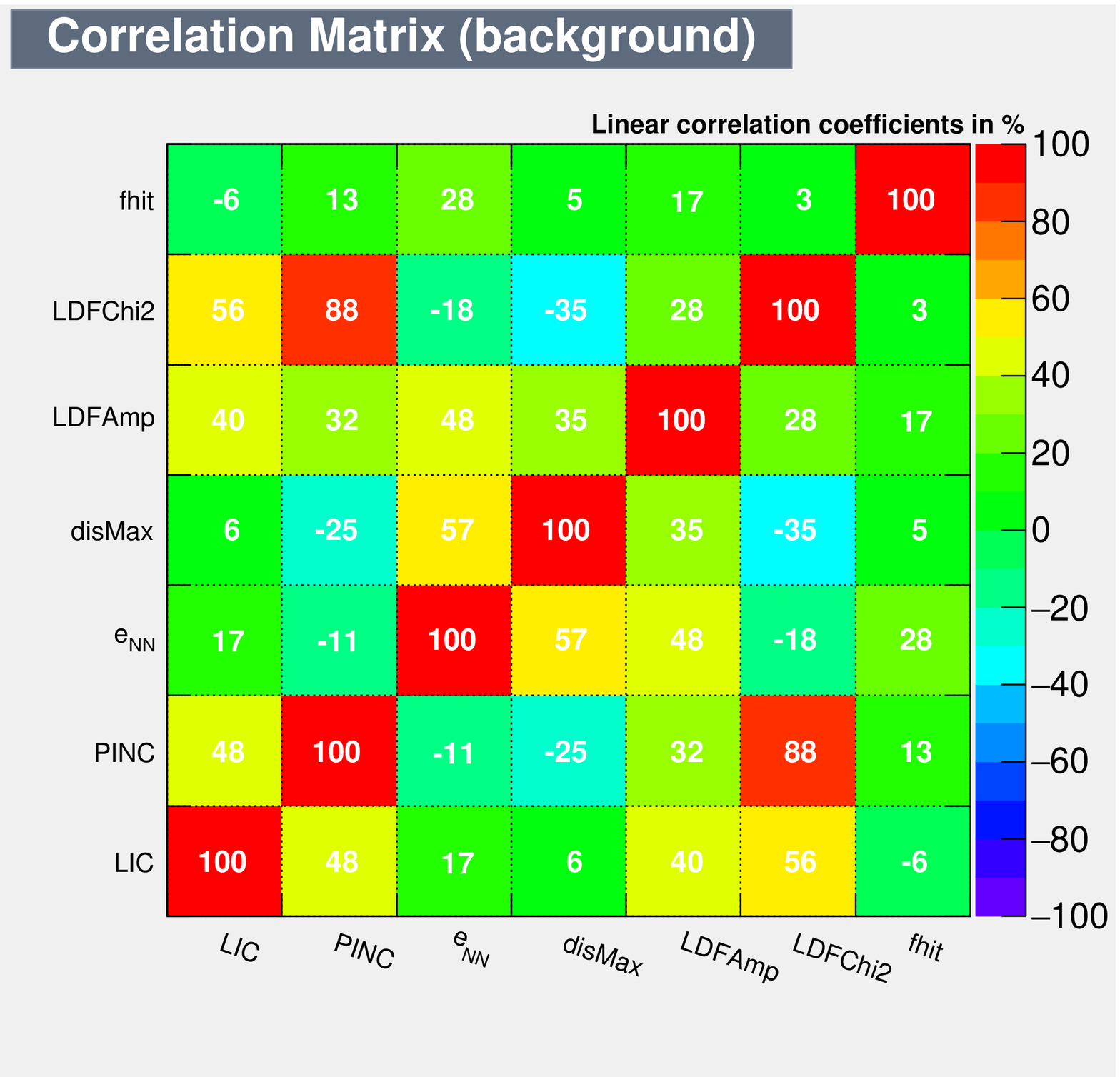}
				\caption{{\small Background}}
%				\label{#7}
			\end{subfigure}
			\caption{{\small The linear correlation matrix for signal (a) and background (b) of each input parameter of the MLT models using MC training set.}}
			\label{fig:corrematrix}
		\end{figure*}

\begin{figure*}[h!]	
			\begin{subfigure}[t]{0.49\textwidth}
				\includegraphics[width=\textwidth,height=0.6\textwidth]{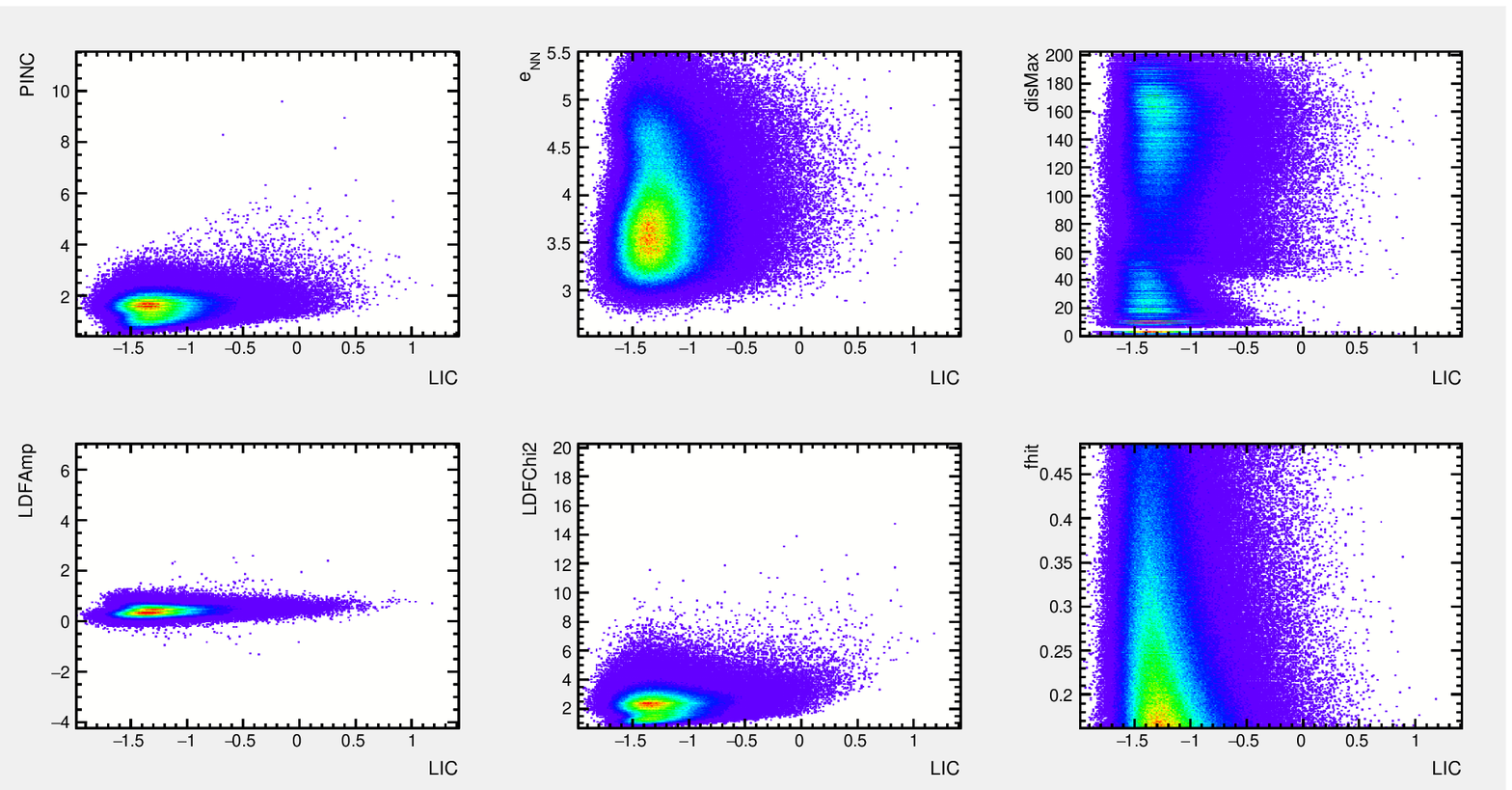}
				\caption{{\small Signal}}
%				\label{#4}
			\end{subfigure}
			\begin{subfigure}[t]{0.49\textwidth}
				\includegraphics[width=\textwidth,height=0.6\textwidth]{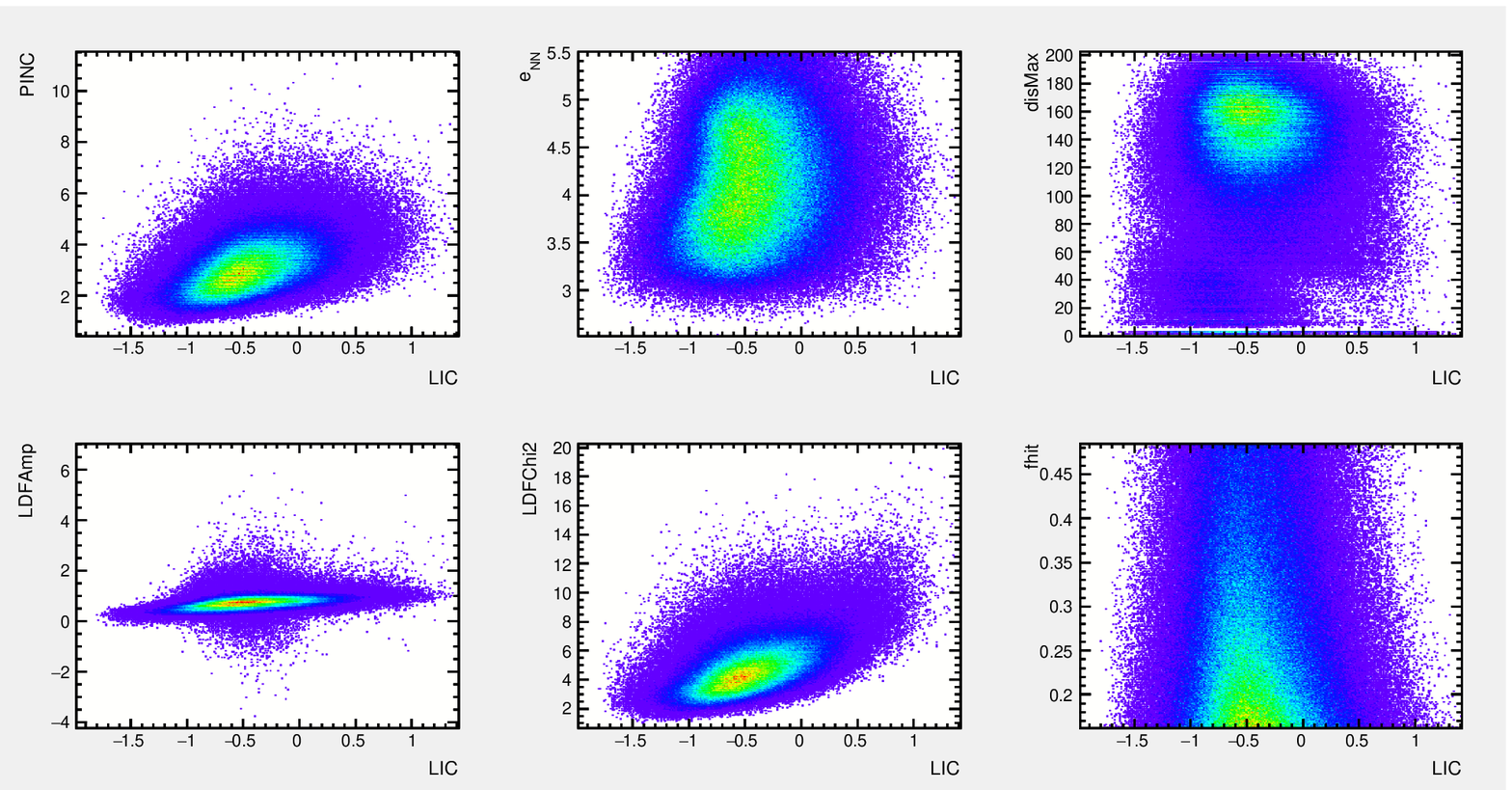}
				\caption{{\small Background}}
%				\label{#7}
			\end{subfigure}
			\caption{{\small The event distribution of two input parameters using simulation training data set for signal and background.}}
			\label{fig:disintputs}
		\end{figure*}

It is considered good practice in MLT to reduce, if possible, the dimensionality (number of input variables) in a model.
One possibility is eliminating one of a pair of heavily correlated variables.
In our simulations, PINC and LDFChi2 are highly correlated in both signal and background (see Figure~\ref{fig:corrematrix}). Figure~\ref{fig:disintputs} shows some of the correlations among variables in MC samples.\\[0.2cm]
%%  correlation matrix 
	
%%  correlation matrix

%%  input histo
	
%%  input histo 
 To test whether the largest correlation was inhibiting ML performance, we trained a BDT after removing PINC; the BDT performance was a few percent worse instead of better. This is consistent with experience in HEP that BDT is often successful using collections of correlated variables.   However, when we trained a NN removing LDFChi2 or PINC, its performance is somewhat worse in some bins and somewhat better in others, and NN generally seemed more sensitive to removal of specific variables than BDT. 
 We would tend to attribute this to the correlations making backpropagation more difficult in NN. BDT optimizes rather differently, by raising weights of mis-classified events to purify leaves.

Table~\ref{tab:ranking} shows the relative importance of the input variables in training on MC data. The NN ordering is based on summed weights applied to the inputs (after linearly normalizing all variables into a range of [-1,1]). The BDT orders variables by the number of times trees use them to define splits. NN and the BDT both rank PINC and LDFChi2 as among the most important variables, but the algorithms appear to use the inputs rather differently, perhaps because NN emphasizes functional dependence, while BDT emphasizes classification more directly. For the High \bbin bin, the BDT ranks \fhit a bit higher than NN does, but it is a low-priority variable for both, at least for MC background training.

%%%%table ranking
\begin{table}[!ht]
		\centering
		\caption{\small Comparison of relative importance of input variables during training using MC background, for NN and BDT.  The variables which are clearly more important are denoted in \textbf{bold}.  The results are shown for each of the 3 trained models, labeled by the $\mathcal{B}$ range covered.}
%		\scalebox{1}[1]{
			\begin{tabular}{c|c|c|c|c|c}
				\hline
                \multicolumn{3}{c|}{NN} & \multicolumn{3}{c}{BDT}\\
                \hline
                $\mathcal{B}$ 0-2& $\mathcal{B}$ 3-5& $\mathcal{B}$ 6-9& $\mathcal{B}$ 0-2& $\mathcal{B}$ 3-5& $\mathcal{B}$ 6-9 \\
				\hline
				\textbf{PINC} & \textbf{PINC} & \textbf{PINC} & 
				\textbf{LDFChi2} & \textbf{PINC} & \textbf{PINC} \\
				\textbf{LDFChi2} & \textbf{LDFChi2} & \textbf{LDFChi2} & 
				\textbf{LiC} & \textbf{LiC} & \textbf{LDFAmp} \\
				\textbf{\fhit} & \textbf{LiC} & \textbf{LDFAmp} & 
				\textbf{PINC} & \textbf{LDFAmp} & \textbf{LDFChi2} \\
				\enn & disMax & \fhit & 
				\textbf{\fhit} & \textbf{LDFChi2} & LiC \\
				LiC & \fhit & disMax & 
				LDFAmp & \fhit & \fhit \\
				disMax & \enn & LiC & 
				\enn & disMax & disMax \\
	    		LDFAmp & LDFAmp & \enn & 
	    		disMax & \enn & \enn\\
				\hline
			\end{tabular}
	    \label{tab:ranking}
%		}
	\end{table}
%%%%table ranking

Differences in correlation effects and variable importance is our best guess as to why difference of the \fhit distribution between real data background and MC signal was interpreted differently by the two ML methods (BDT seemed to ignore this difference, but NN lost performance). Using MC for both background and signal had the virtue of consistent energy distributions and \fhit (PMT availability), and in fact demonstrated improvements over the SC trained on MC signal and real data background. However, using \fhit and \enn as interpolation variables may have made ML methods more vulnerable compared to SC.
		
%================================================ agradecimientos =========================================================================
%https://private.hawc-observatory.org/wiki/images/e/e3/HAWCAcknowledgement30Jun21LaTEX.txt
\medskip

{\bf Acknowledgments} We acknowledge the support from: the US National Science Foundation (NSF); the US Department of Energy Office of High-Energy Physics; the Laboratory Directed Research and Development (LDRD) program of Los Alamos National Laboratory; Consejo Nacional de Ciencia y Tecnolog\'ia (CONACyT), M\'exico, grants 271051, 232656, 260378, 179588, 254964, 258865, 243290, 132197, A1-S-46288, A1-S-22784, c\'atedras 873, 1563, 341, 323, Red HAWC, M\'exico; DGAPA-UNAM grants IG101320, IN111716-3, IN111419, IA102019, IN110621, IN110521; VIEP-BUAP; PIFI 2012, 2013, PROFOCIE 2014, 2015; the University of Wisconsin Alumni Research Foundation; the Institute of Geophysics, Planetary Physics, and Signatures at Los Alamos National Laboratory;~~Polish Science Centre grant, DEC-2017/27/B/ST9/02272; Coordinaci\'on de la Investigaci\'on Cient\'ifica de la Universidad Michoacana; Royal Society - Newton Advanced Fellowship 180385; Generalitat Valenciana, grant CIDEGENT/2018/034; Chulalongkorn University’s CUniverse (CUAASC) grant; Coordinaci\'on General Acad\'emica e Innovaci\'on (CGAI-UdeG), PRODEP-SEP UDG-CA-499; Institute of Cosmic Ray Research (ICRR), University of Tokyo, H.F. acknowledges support by NASA under award number 80GSFC21M0002; This research was partially carried out using the HKU Information Technology Services research computing facilities that are supported in part by the Hong Kong UGC Special Equipment Grant (SEG HKU09). P.S.P. and T.C. were supported at HKU by a grant from the Big Data Project Fund (BDPF) and a GRF grant (Project 17304920) from the Hong Kong Government. We also acknowledge the significant contributions over many years of Stefan Westerhoff, Gaurang Yodh and Arnulfo Zepeda Dominguez, all deceased members of the HAWC collaboration. Thanks to Scott Delay, Luciano D\'iaz and Eduardo Murrieta for technical support.

  \bibliographystyle{elsarticle-num} 
  \bibliography{biblio}

\end{document}